\documentclass[
    aip, jcp, amsmath, amssymb, reprint, floatfix]{revtex4-2}
\usepackage{amsmath,amssymb}
\usepackage{graphicx}
\makeatletter
\providecommand{\@captype}{figure}
\makeatother

\begin{document}

\title{Tersoff-NN: Hybrid Machine-Learned Bond-Order Extension of the Tersoff Potential} %Title of paper

\author{Yusuke Nishimura}
\email{yusukeskelton@toki.waseda.jp}
\author{Sho Kubota}
\author{Kotaro Takematsu}
\author{Kentaro Hirai}
\author{Takanobu Watanabe}
\affiliation{Faculty of Science and Engineering, Waseda University, Shinjuku, Tokyo 169-8555, Japan}

\date{October 11, 2026}

\begin{abstract}
Interatomic potentials combining accuracy and robustness with computational efficiency are highly desirable for large-scale molecular dynamics simulations. Although machine-learning interatomic potentials can reproduce first-principles potential-energy surfaces with high accuracy, they typically incur higher computational costs than conventional empirical potentials and may be less reliable outside their training domains. To address this limitation, we developed Tersoff-NN---a hybrid machine-learning potential for Si, which retains the analytical radial interactions and cutoff structure of the Tersoff potential while representing its environment-dependent bond-order factor using compact neural networks (NNs). Benchmarks for 32,768-atom systems show that Tersoff-NN is 4--12 times faster than the Behler--Parrinello-type NN potentials tested in this study. Tersoff-NN improves the short-range structure of liquid Si relative to the conventional Tersoff potential, closely reproduces the ab initio radial and angular distributions, and reproduces the equilibrium structure of diamond Si and qualitative energetic trends among several crystalline phases. Furthermore, melt-quench simulations reveal a disordered solid with predominantly four-coordinated, tetrahedral-like local environments; additionally, characteristic reconstruction motifs are retained on Si surfaces despite their exclusion from the training dataset. In addition to these structural properties, a direct comparison with an alternative atom-centered bond-order NN highlights the critical role of physical constraints in ensuring dynamic stability. Despite exhibiting comparable validation errors, the selected triplet-based architecture remains stable, while an alternative model exhibits unphysical bond-order growth and atomic collapse. These results confirm that learning only the bond-order term improves the description of diverse Si environments while retaining substantial computational efficiency and enhancing dynamical stability.
\end{abstract}

\maketitle
\section{Introduction}

The accuracy and practicality of molecular dynamics (MD) simulations are largely governed by the interatomic potential used to compute atomic forces. When developing or selecting an interatomic potential, the relative importance of accuracy, applicability, and computational cost must be considered based on the target application. Classical interatomic potentials (IAPs) are computationally efficient and can be applied to large-scale and long-time MD simulations; however, their rigid functional forms and system-specific parameterizations can limit their transferability across structurally distinct environments. In contrast, machine learning (ML) IAPs (ML-IAPs) \cite{zuoPerformanceCostAssessment2020} trained on ab initio datasets have achieved accuracy approaching that of density functional theory (DFT) across diverse atomic configurations; however, this achievement typically comes at the expense of increased computational cost.

Recent advancements in ML-IAPs include development of local-environment-based frameworks such as Behler--Parrinello-type high-dimensional neural network (NN) potentials \cite{behlerGeneralizedNeuralNetworkRepresentation2007,behlerAtomcenteredSymmetryFunctions2011,singraberLibraryBasedLAMMPSImplementation2019}, Gaussian approximation potentials\cite{bartokGaussianApproximationPotentials2010,bartokRepresentingChemicalEnvironments2013,bartokMachineLearningGeneralPurpose2018}, spectral neighbor analysis potentials \cite{thompsonSpectralNeighborAnalysis2015a}, moment tensor potentials\cite{shapeevMomentTensorPotentials2016}, and deep potential models\cite{zhangDeepPotentialMolecular2018} as well as graph-NN and equivariant architectures such as NequIP\cite{batznerE3equivariantGraphNeural2022} and message-passing atomic cluster expansion\cite{batatiaMACEHigherOrder2022}. Although these models have substantially improved the accuracy and applicability of atomistic simulations, this improvement is often accompanied by increased model complexity and computational overhead. Systematic benchmarks of representative ML-IAPs\cite{zuoPerformanceCostAssessment2020,leimerothMachinelearningInteratomicPotentials2025} have demonstrated that a higher accuracy is often achieved at the expense of higher model complexity and computational cost, revealing a broad accuracy--speed tradeoff among modern ML-IAPs. Therefore, despite the remarkable accuracy of ML-IAPs, their computational cost remains a practical bottleneck for long-term and large-scale MD simulations.

These considerations have motivated development of hybrid and physics-informed ML-IAPs, wherein ML is embedded into selected components of an established analytical potential rather than being used to replace the entire potential-energy surface. These designs aim to combine the flexibility of ML with the interpretability and computational efficiency of physically motivated functional forms. Representative examples include ReaxFF-MPNN and ReaxFF-nn\cite{xueReaxFFMPNNMachineLearning2021,yeReaxFFnnReactiveMachinelearning2025}, wherein NNs are introduced into the bond-order-related part of a reactive force field while retaining other ReaxFF energy terms.

Furthermore, analytical bond-order models offer a natural basis for such hybrid approaches by separating radial interactions from the environmental dependence of bond strength. The Tersoff potential represents a compact realization of this concept, rooted in Abell's chemical pseudopotential theory \cite{abellEmpiricalChemicalPseudopotential1985,tersoffNewEmpiricalApproach1988}. Its bond-order factor incorporates the effects of coordination and local geometry, allowing different local bonding environments to be described within a common framework. This structure allows ML to refine environment-dependent bond strength while preserving the underlying analytical framework. Early examples include NN replacement of the many-body bond-order term in Brenner--Tersoff-type potentials\cite{hobdayApplicationsNeuralNetworks1999} and NN-assisted parameterization of analytical Tersoff potentials\cite{malsheParametrizationAnalyticInteratomic2008}. More recent advancements include development of physically informed NN potentials based on analytical bond-order models\cite{punPhysicallyInformedArtificial2019} and machine-learned bond-order potentials learning the environmental dependence of bond order while retaining classical two-body terms\cite{kohataMachineLearnedBondOrderPotential2025}.

Although these studies offer means for combining ML with physically motivated functional forms, they did not establish that retaining an analytical framework offers a practical computational advantage. The reported implementations exhibited additional computational overhead relative to either the original analytical potential or conventional NN potentials\cite{hobdayApplicationsNeuralNetworks1999,punPhysicallyInformedArtificial2019}. Therefore, it remains unclear whether a Tersoff-type hybrid model can improve accuracy while retaining a substantial speed advantage over conventional ML-IAPs.

This study developed Tersoff-NN---a lightweight hybrid ML-IAP based on the Tersoff bond-order potential for Si\cite{tersoffNewEmpiricalApproach1988,tersoffEmpiricalInteratomicPotential1988,tersoffModelingSolidstateChemistry1989}. The analytical radial interactions and cutoff function were retained, while the environment-dependent bond-order factor was represented by compact NNs. The analytical parameters were determined before NN training and fixed subsequently, confining ML to the local environment dependence of bond strength. The model was trained on ab initio energies, forces, and stresses for Si configurations in crystalline and disordered environments. This study addressed the key question of whether learning only the environment-dependent bond-order response is sufficient to improve description of diverse Si environments while retaining a practical computational advantage. We evaluated description of crystalline energetics, elastic and vibrational properties, finite-temperature dynamics, liquid structure, and melt-quenched disordered Si. The out-of-domain behavior was further examined using reconstructed Si surfaces excluded from the training dataset. In addition, large-scale MD benchmarks and a component-level timing analysis were employed to validate the practical computational advantage offered by the proposed restricted learning strategy over conventional descriptor-based NN potentials and to identify the origin of the observed performance difference.

The remainder of this paper is organized as follows. Section~II presents the formulation and training strategy of Tersoff-NN. Section~III evaluates the performance of the proposed architecture in terms of fitting accuracy, dynamical stability, and computational efficiency, and its behavior across diverse Si environments. Section~IV summarizes the main findings and conclusions of this study.

\section{Methods}

\subsection{Formulation of Tersoff-NN}

Tersoff-NN is designed based on a bond-order representation developed by Abell\cite{abellEmpiricalChemicalPseudopotential1985} and formulated in a practical form by Tersoff\cite{tersoffEmpiricalInteratomicPotential1988,tersoffModelingSolidstateChemistry1989}, as discussed in the Introduction. In the Tersoff form, energy assigned to each bond is defined as the sum of a repulsive pair term and a bond-order-modulated attractive pair term. The bond-order factor adjusts the magnitude of the attractive contribution based on the local atomic environment, thereby representing changes in the bond strength.

Based on this separation, Tersoff-NN retains the analytical two-body terms and introduces NN flexibility only into the bond-order term. This enables the model to improve description of local many-body environments while preserving the interpretable and lightweight analytical backbone of the original Tersoff potential.

\begin{figure*}[tbp]
\centering
\begin{minipage}{0.84\textwidth}
\raggedright
\textbf{(a)}\\
\centering
\includegraphics[width=\linewidth]{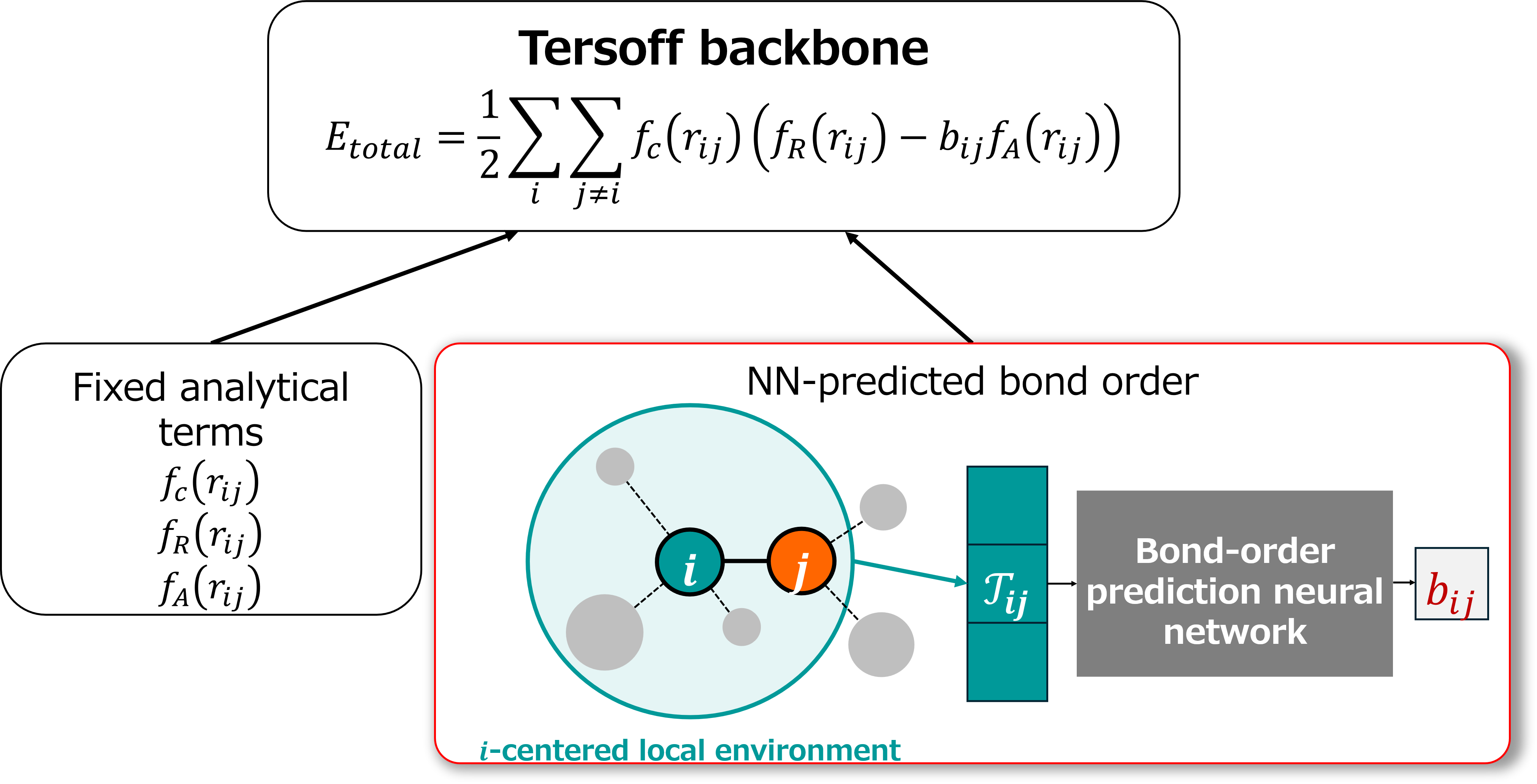}
\end{minipage}

\vspace{0.5em}

\begin{minipage}[t]{0.45\textwidth}
\raggedright
\textbf{(b)}\\
\centering
\includegraphics[width=\linewidth]{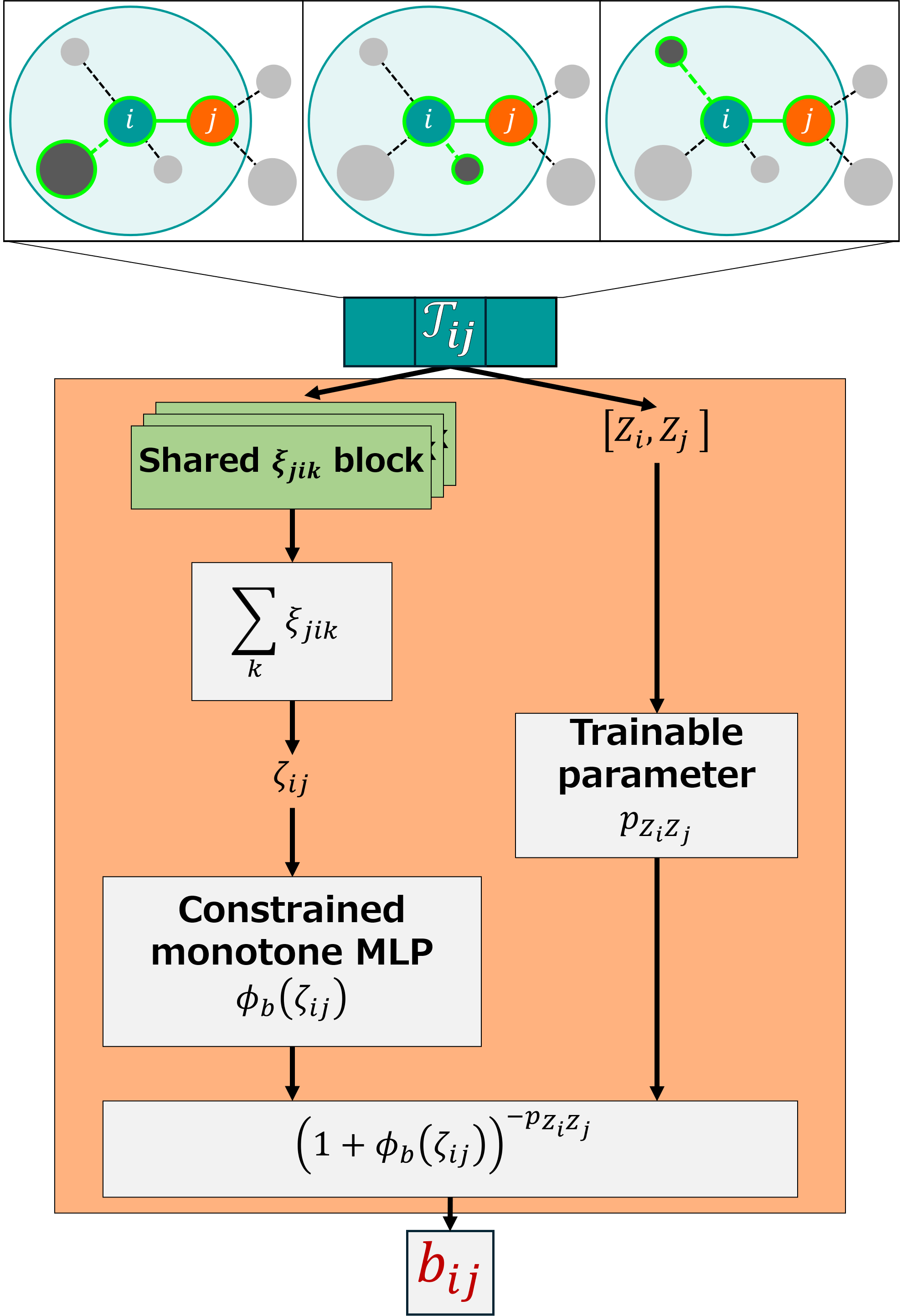}
\end{minipage}
\hspace{0.04\textwidth}
\begin{minipage}[t]{0.315\textwidth}
\raggedright
\textbf{(c)}\\
\centering
\includegraphics[width=\linewidth]{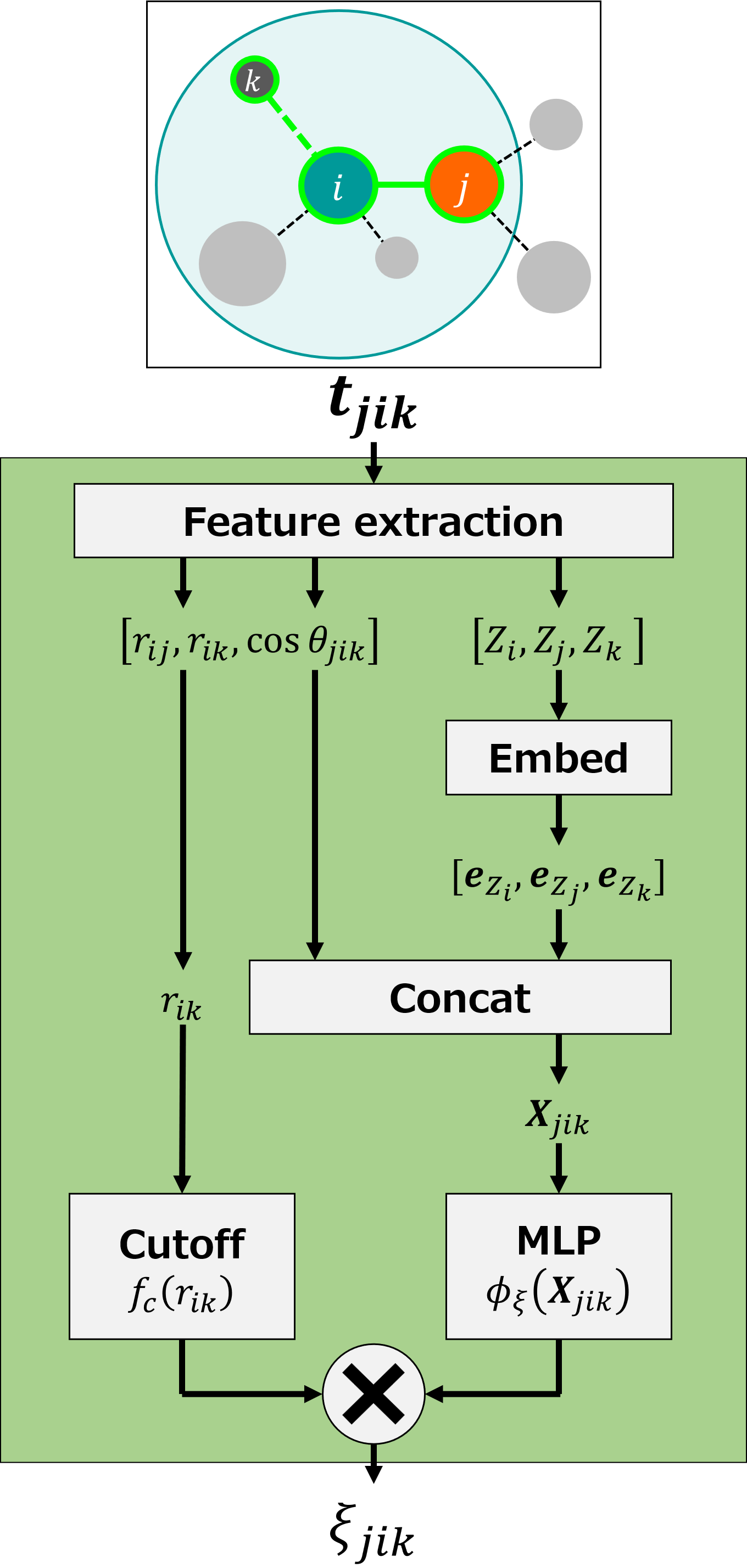}
\end{minipage}
\caption{Detailed architecture of Tersoff-NN with triplet learned-map BOPNN. (a) Analytical cutoff function and repulsive and attractive pair terms of the Tersoff potential are retained, and the environment-dependent bond-order factor $b_{ij}$ is predicted using an NN. (b) For each bond $i$--$j$, contributions from neighboring atoms $k$ are aggregated and combined with species-dependent trainable parameters to construct $b_{ij}$. (c) Each triplet contribution is obtained from geometric and species features of the local environment.}
\label{fig:model-structure}
\end{figure*}

The proposed modeling strategy aligns closely with a physics-informed framework developed by Kohata \emph{et al.}\cite{kohataMachineLearnedBondOrderPotential2025} for a machine-learned bond-order potential, which combined analytical radial interactions with representation of environment-dependent bond-order effects using NNs. Their formulation emphasizes compactness and interpretability while retaining sufficient flexibility to describe diverse local environments. Meanwhile, Tersoff-NN adopts a deliberately more constrained formulation: the conventional Tersoff radial terms are retained, and ML is confined to the attractive bond-order factor. We refer to the NN component used to predict the environment-dependent bond-order factor \(b_{ij}\) as the bond-order prediction neural network (BOPNN). This restricted architecture is intended to preserve the compact computational structure of the Tersoff potential while introducing flexibility into its principal many-body component. Several BOPNN architectures were considered in this study, as discussed in the following subsection. Fig.~\ref{fig:model-structure} illustrates the Tersoff-NN architecture using the triplet learned-map BOPNN, with panel (a) displaying the overall Tersoff backbone and BOPNN structure.

The total energy in Tersoff-NN is expressed as
\begin{equation}
E_{\mathrm{total}} = \frac{1}{2}\sum_{i}\sum_{j \neq i} f_{c}\left( r_{ij} \right)\left[ f_{R}\left( r_{ij} \right) - b_{ij}f_{A}\left( r_{ij} \right) \right],
\end{equation}
where \(r_{ij}\), \(f_{c}\), \(f_{R}\), \(f_{A}\), and \(b_{ij}\) represent the distance between atoms \(i\) and \(j\), a smooth cutoff function, the repulsive pair term, the attractive pair term, and the bond-order factor, respectively. The repulsive and attractive terms are expressed as
\begin{equation}
f_{R}\left( r_{ij} \right) = A_{ij}\exp\left( - \lambda_{1}r_{ij} \right),
\end{equation}
\begin{equation}
f_{A}\left( r_{ij} \right) = B_{ij}\exp\left( - \lambda_{2}r_{ij} \right),
\end{equation}
where \(A_{ij}\), \(B_{ij}\), \(\lambda_{1}\), and \(\lambda_{2}\) are the analytical two-body parameters determined before training the NN model. For the present Si model, \(R=3.25\)~\AA\ and \(D=0.25\)~\AA, corresponding to a smooth cutoff function transition from 3.0 to 3.5~\AA. The smooth cutoff function is defined as
\begin{equation}
f_{c}\left( r_{ij} \right) =
\begin{cases}
1, & r_{ij} \le R - D, \\
\tfrac{1}{2} - \tfrac{1}{2}\sin\!\left( \frac{\pi\left( r_{ij} - R \right)}{2D} \right), & \left| r_{ij} - R \right| < D, \\
0, & r_{ij} \ge R + D.
\end{cases}
\end{equation}

\subsection{Architectural variants of BOPNN}

\begin{figure}[tbp]
\centering
\makebox[\linewidth][l]{\textbf{(a)}}\\
\includegraphics[width=0.95\linewidth]{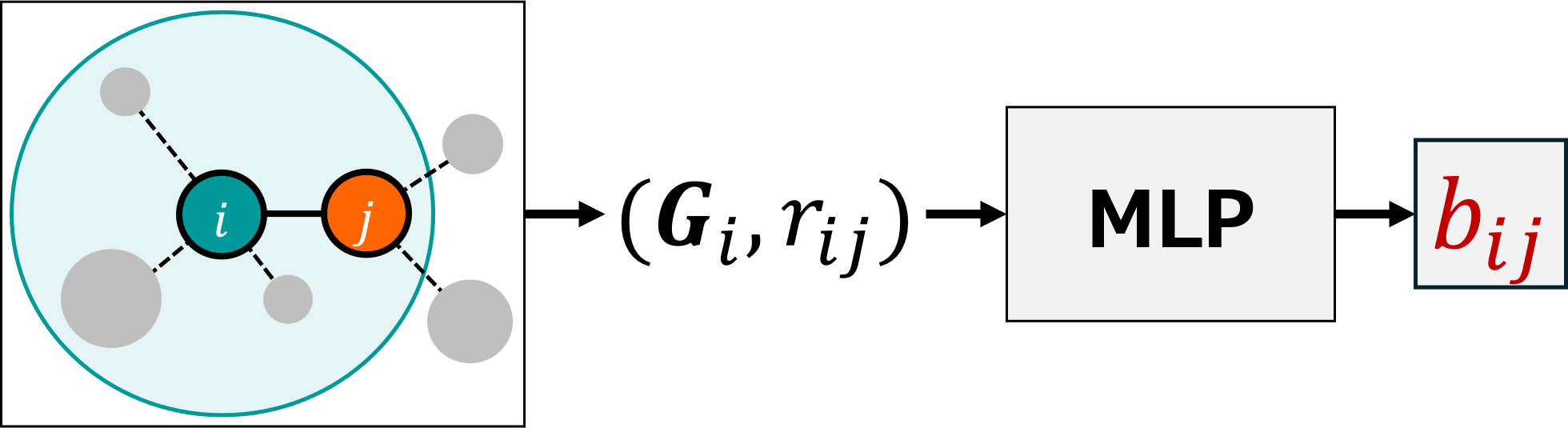}
\par\vspace{0.4em}
\makebox[\linewidth][l]{\textbf{(b)}}\\
\includegraphics[width=0.95\linewidth]{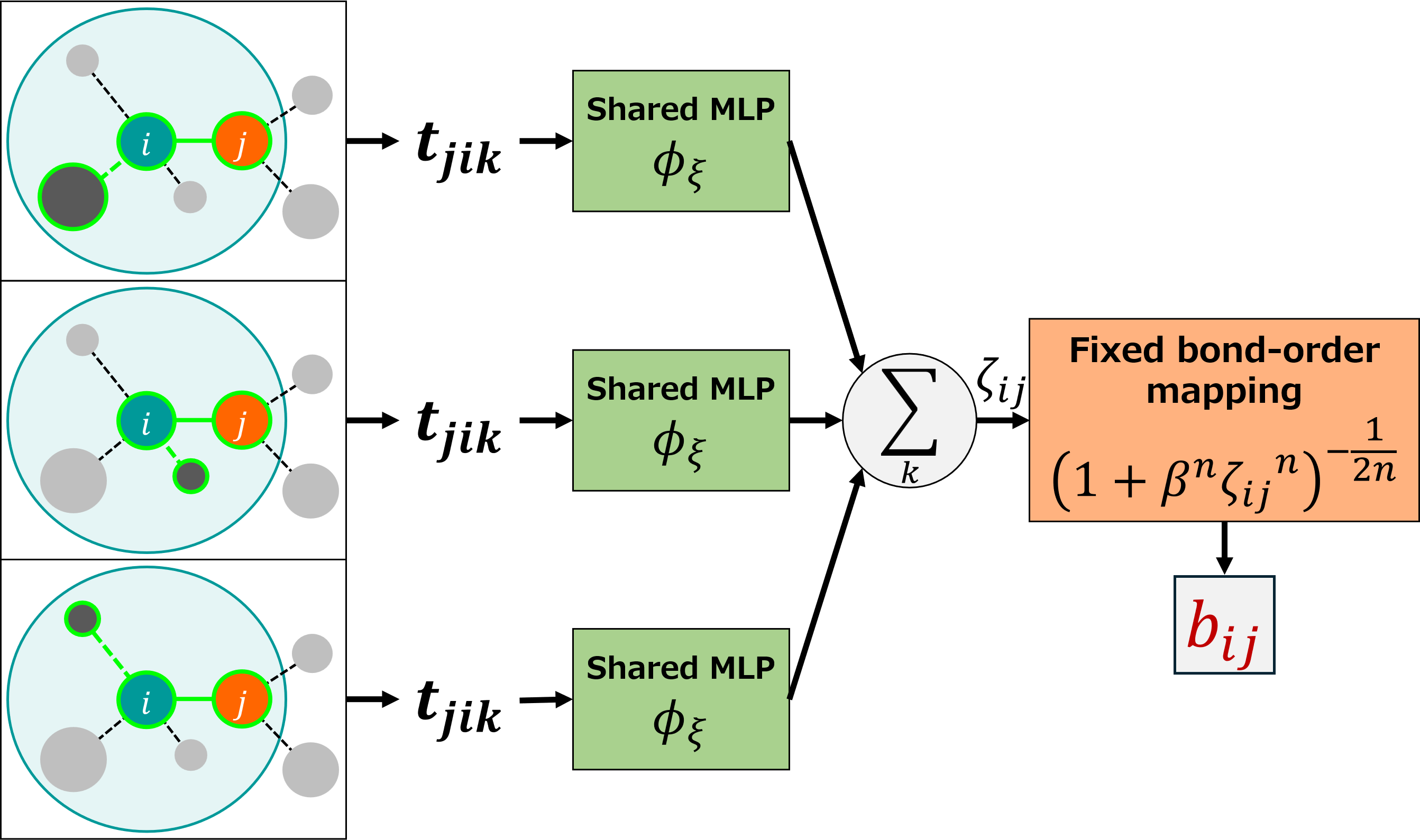}
\par\vspace{0.4em}
\makebox[\linewidth][l]{\textbf{(c)}}\\
\includegraphics[width=0.95\linewidth]{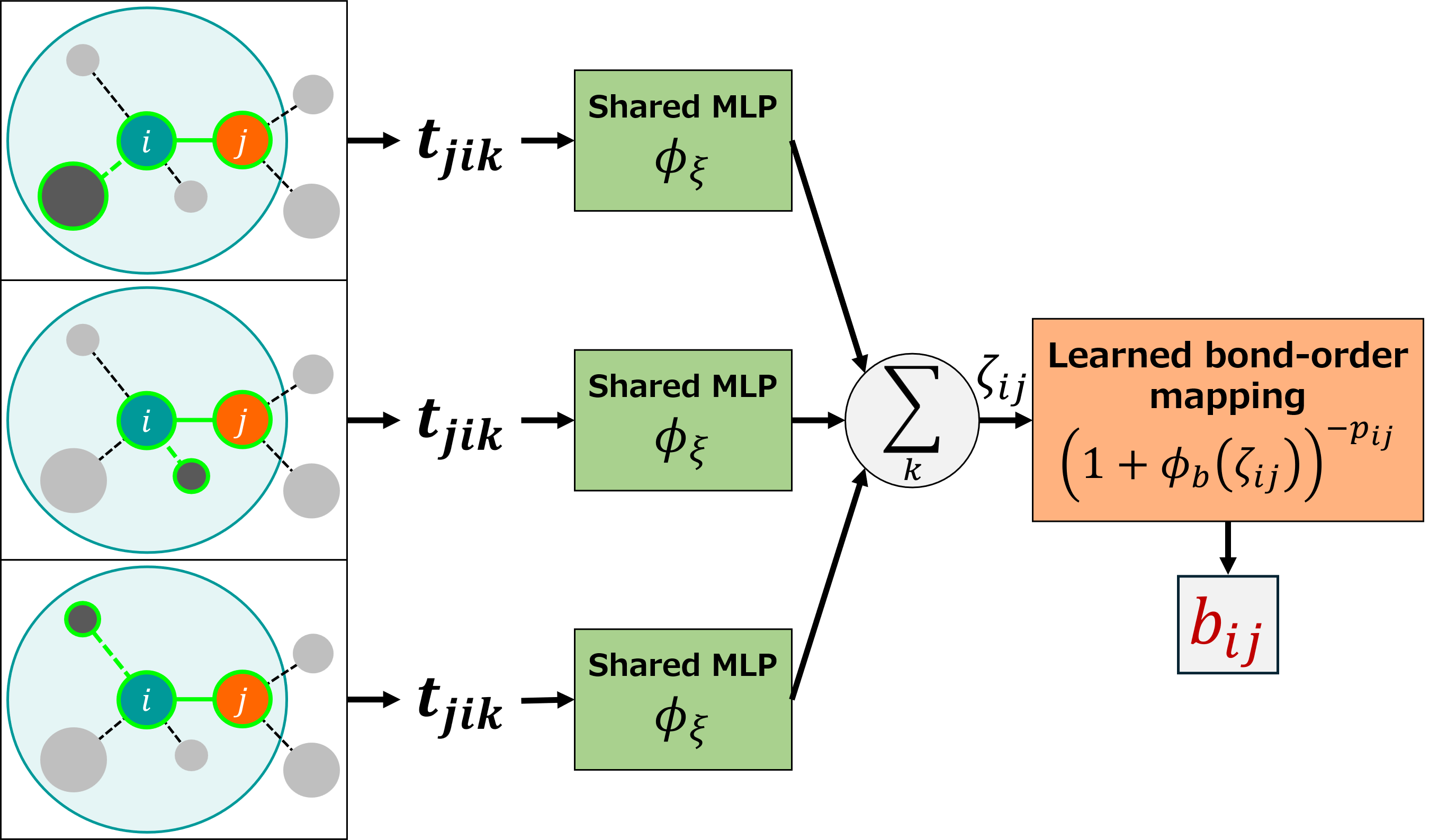}
\caption{Architectural variants of BOPNN. (a)
ACSF-BOPNN directly predicts the bond-order factor \(b_{ij}\) from an atom-centered symmetry-function vector \(\mathbf{G}_{i}\) and the selected pair distance \(r_{ij}\). (b) The triplet fixed-map BOPNN aggregates learned triplet contributions into the effective coordination variable \(\zeta_{ij}\) and applies the analytical Tersoff bond-order mapping. (c) The triplet learned-map BOPNN replaces the analytical mapping with a constrained NN mapping. For compactness, \(t_{jik}\) denotes the triplet \(j-i-k\), represented by the feature vector \(\mathbf{X}_{jik}\) defined in Eq.~\eqref{eq:triplet-feature}.}
\label{fig:bopnn-variants}
\end{figure}

We considered three architectural variants of the BOPNN to examine the roles of local-environment representation and post-aggregation bond-order mapping. The former concerns how information about neighboring atoms is encoded prior to aggregation, whereas the latter concerns how the resulting aggregated quantity is converted into the bond-order value. All BOPNN variants retain the same analytical Tersoff radial backbone and predict the environment-dependent bond-order factor \(b_{ij}\); however, they differ in the representation of the local environment and mapping of environmental information onto \(b_{ij}\). Fig.~\ref{fig:bopnn-variants} summarizes the three BOPNN architectures considered in this study.

The first variant---denoted as atom-centered symmetry functions BOPNN (ACSF-BOPNN)---directly predicts the bond-order factor from an atom-centered symmetry-function vector and the distance of the selected bond as follows:
\begin{equation}
b_{ij}^{\mathrm{ACSF}}=\phi_{\mathrm{ACSF}}\left(\mathbf{G}_{i}, r_{ij}\right),
\end{equation}
where \(\mathbf{G}_{i}\) and \(\phi_{\mathrm{ACSF}}\) represent a vector of radial and angular ACSFs \cite{behlerAtomcenteredSymmetryFunctions2011} and a multilayer perceptron, respectively. The same atom-centered descriptor \(\mathbf{G}_{i}\) was used for all ordered bonds originating from atom \(i\), while the selected neighbor \(j\) entered the model explicitly only through the pair distance \(r_{ij}\). The definitions and parameter sets of ACSFs are provided in the Supplementary Material.

The second variant---denoted as the triplet fixed-map BOPNN---introduces a bond-conditioned triplet representation. For each ordered bond \(i - j\), neighboring atoms \(k\) around the central atom \(i\) are considered, with \(k \neq j\). For each triplet \(j - i - k\), the invariant geometric feature vector is defined as
\begin{equation}
\mathbf{X}_{jik} = \left\lbrack r_{ij},r_{ik},\cos\theta_{jik},\mathbf{e}_{Z_{i}},\mathbf{e}_{Z_{j}},\mathbf{e}_{Z_{k}} \right\rbrack,
\label{eq:triplet-feature}
\end{equation}
where \(\theta_{jik}\) represents the angle formed by bonds \(i - j\) and \(i - k\) and \(\mathbf{e}_{Z_{i}}\), \(\mathbf{e}_{Z_{j}}\), and \(\mathbf{e}_{Z_{k}}\) are the learnable embeddings of atomic species \mbox{\cite{schuttQuantumchemicalInsightsDeep2017,schuttSchNetDeepLearning2018}}. The embedding vector is bounded element-wise by a sigmoid function
\begin{equation}
\mathbf{e}_{Z} = \varsigma\left( \mathbf{W}_{emb}\lbrack Z\rbrack \right),
\end{equation}
where \(\mathbf{W}_{emb}\) and \(\varsigma\) represent the embedding table and sigmoid function, respectively. Although the present study considered a unary Si system, the embedding layer was retained for future extensions to multicomponent systems.

The triplet contribution \(\xi_{jik}\) is subsequently calculated as
\begin{equation}
\xi_{jik} = f_{c}\left( r_{ik} \right)\mathrm{Softplus}\left( \phi_{\xi}\left( \mathbf{X}_{jik} \right) \right),
\end{equation}
where \(\phi_{\xi}\) represents a multilayer perceptron and
\(\mathrm{Softplus}(x)=\ln(1+\exp(x))\) \cite{dugasIncorporatingSecondorderFunctional2000}. The factor \(f_{c}\left(r_{ik}\right)\) ensures that the triplet contribution vanishes smoothly at the cutoff distance. Meanwhile, Softplus output activation and the non-negative cutoff function guarantee that \(\xi_{jik} \geq 0\).

Let $\mathcal{N}_{i}$ represent the set of neighboring atoms of atom $i$ within the cutoff radius. The effective coordination variable for the bond \(i - j\) is obtained by summing the triplet contributions over neighboring atoms
\begin{equation}
\zeta_{ij} = \sum_{\substack{k\in\mathcal{N}_{i}\\k\neq j}}\xi_{jik}.
\end{equation}
The resulting effective coordination variable is converted into the bond-order factor via the analytical Tersoff mapping.
\begin{equation}
b_{ij}^{\mathrm{fixed}}=\left(1+\beta_{ij}^{\,n_{ij}}\zeta_{ij}^{\,n_{ij}}\right)^{-\frac{1}{2n_{ij}}},
\end{equation}
where \(\beta_{ij}\) and \(n_{ij}\) represent the parameters of the analytical bond-order function. This variant was introduced to assess the effect of retaining a fixed analytical relationship between the aggregated environmental variable and the bond-order factor.

The third variant---the triplet learned-map BOPNN---uses the same triplet representation, triplet network \(\phi_{\xi}\), and aggregation procedure as the triplet fixed-map BOPNN. However, the analytical mapping from \(\zeta_{ij}\) to \(b_{ij}\) is replaced by a constrained NN. The final bond-order factor is computed as
\begin{equation}
b_{ij} = \left( 1 + \phi_{b}\left( \zeta_{ij} \right) \right)^{- p_{ij}},
\end{equation}
where \(\phi_{b}\) and \(p_{ij}\) represent a monotonic NN and a trainable species-pair-dependent exponent, respectively. The monotonic NN is constructed using non-negative weights \cite{danielsMonotonePartiallyMonotone2010,runjeConstrainedMonotonicNeural2023} without any bias terms. The exponent \(p_{ij}\) is constrained to remain positive during training. The monotonic network uses a shifted Softplus activation \(\mathrm{SSP}(x)=\ln(1+\exp(x))-\ln 2\) \cite{schuttSchNetDeepLearning2018}. Because the weights are non-negative and SSP is monotonically increasing, \(\phi_b\) is monotonically increasing with respect to \(\zeta_{ij}\). Furthermore, the absence of bias terms and \(\mathrm{SSP}(0)=0\) ensure that \(\phi_b(0)=0\) and \(\phi_b(\zeta_{ij})\geq0\) for \(\zeta_{ij}\geq0\). This construction guarantees \(b_{ij} = 1\) when \(\zeta_{ij} = 0\), corresponding to an isolated dimer, and causes \(b_{ij}\) to decrease monotonically with an increase in effective coordination. Therefore, the NN preserves the qualitative behavior of the original Tersoff bond-order expression while enabling a more flexible dependence on local three-body environments.

A comparison between the ACSF-BOPNN and triplet-based variants was carried out to evaluate the atom-centered and bond-conditioned architectures. The comparison between the two triplet-based variants was designed to assess the role of learnable flexibility after aggregating individual triplet contributions.

For architectural comparison, the ACSF-BOPNN used a 23-dimensional input vector comprising 22 atom-centered symmetry functions and the selected pair distance \(r_{ij}\). Its multilayer perceptron included two hidden layers with 32 neurons each. Both the triplet-based architectures used the same 12-dimensional triplet feature vector and the same triplet network \(\phi_{\xi}\). In the triplet fixed-map BOPNN, the bond-order mapping parameters \(\beta_{ij}\) and \(n_{ij}\) were fixed to the values of the Si parameter set reported by Tersoff\cite{tersoffEmpiricalInteratomicPotential1988}; additionally, these parameter values were not optimized during training. Furthermore, the triplet learned-map BOPNN used the bond-order network \(\phi_{b}\) with two hidden layers of eight neurons each to transform the scalar effective coordination variable \(\zeta_{ij}\) with the trainable exponent \(p_{ij}\). The hidden layers of the unconstrained networks used a sigmoid linear unit (SiLU) activation function\cite{elfwingSigmoidweightedLinearUnits2018}. These network sizes were selected to provide the three variants with comparable model capacities while retaining the compact structure of the bond-order model.

\begin{table}[tbp]
\caption{Network architectures and trainable parameter counts of
BOPNN variants. Parameter counts include all trainable network weights and biases, species embeddings, and trainable exponent \(p_{ij}\), where applicable. Fixed analytical parameters including \(\beta_{ij}\) and \(n_{ij}\) are excluded.}
\label{tab:bopnn-architectures}
\centering
\small
\setlength{\tabcolsep}{2pt}
\begin{tabular}{@{}cccc@{}}
\hline\hline
Model & Main network &
\begin{tabular}[c]{@{}c@{}}Post-aggregation\\map\end{tabular} &
\(N_{\mathrm{param}}\) \\
\hline
ACSF-BOPNN & \(23\)-\(32\)-\(32\)-\(1\) & Direct prediction & 1857 \\
\shortstack{Triplet fixed-map\\BOPNN} & \(12\)-\(32\)-\(32\)-\(1\) & Analytical & 1508 \\
\shortstack{Triplet learned-map\\BOPNN} & \(12\)-\(32\)-\(32\)-\(1\) & \(1\)-\(8\)-\(8\)-\(1\) & 1589 \\
\hline\hline
\end{tabular}
\end{table}

\subsection{Parameter optimization and model training}

The analytical radial and cutoff parameters of Tersoff-NN were determined prior to training the NN bond-order model. In this framework, the analytical parameters were initially adjusted to reproduce the energetics of a selected reference phase and subsequently fixed, and NN optimization was confined to an environment-dependent bond-order factor.

For the present Si model, diamond Si was selected as the reference crystalline phase. After selecting an extended cutoff range, the radial and cutoff parameters were fitted to its DFT cohesive energy--volume relationship, while the original Tersoff bond-order expression was fixed. After this preliminary optimization, the analytical parameters were fixed, and only the bond-order factor \(b_{ij}\) was optimized using the NN model. The optimized radial and cutoff parameters and details of the fitting procedure are provided in the Supplementary Material.

The NN bond-order model was trained using a DFT-labeled Si dataset covering various local environments---from crystalline diamond structures to liquid, amorphous, interfacial, and defect-containing configurations. This dataset was designed to sample both near-equilibrium and strongly distorted finite-temperature bonding environments. Details of the dataset composition and generation procedures are provided in the Supplementary Material.

DFT reference calculations were performed using Vienna ab initio simulation package \cite{kresseEfficientIterativeSchemes1996,kresseUltrasoftPseudopotentialsProjector1999}, utilizing the projector augmented-wave method and Perdew—Burke--Ernzerhof (PBE) exchange-correlation functional \cite{perdewGeneralizedGradientApproximation1996}. The generated structures and DFT labels were processed using Atomic Simulation Environment (ASE) \cite{hjorthlarsenAtomicSimulationEnvironment2017}. The final dataset was randomly divided into training and validation subsets at an 80:20 ratio. A total of 16,874 configurations was stratified by structural category and split at the configuration level into 13,499 training and 3,375 validation configurations using a fixed seed. The NN model was trained on DFT energies, forces, and stresses using a weighted loss function. All three BOPNN variants were trained for 1000 epochs using Adam at an initial learning rate of \(10^{-3}\) and a batch size of 64. Detailed hyperparameters and loss weights are provided in the Supplementary Material. Input geometric features were scaled using statistics from the training subset, and the same scaling parameters were applied to the validation subset. The training hyperparameters, loss weights, and category-wise validation errors are provided in the Supplementary Material.

\section{Results and Discussion}

\subsection{Comparison of BOPNN architectures}

\begin{figure}[tbp]

\centering
\includegraphics[width=\linewidth]{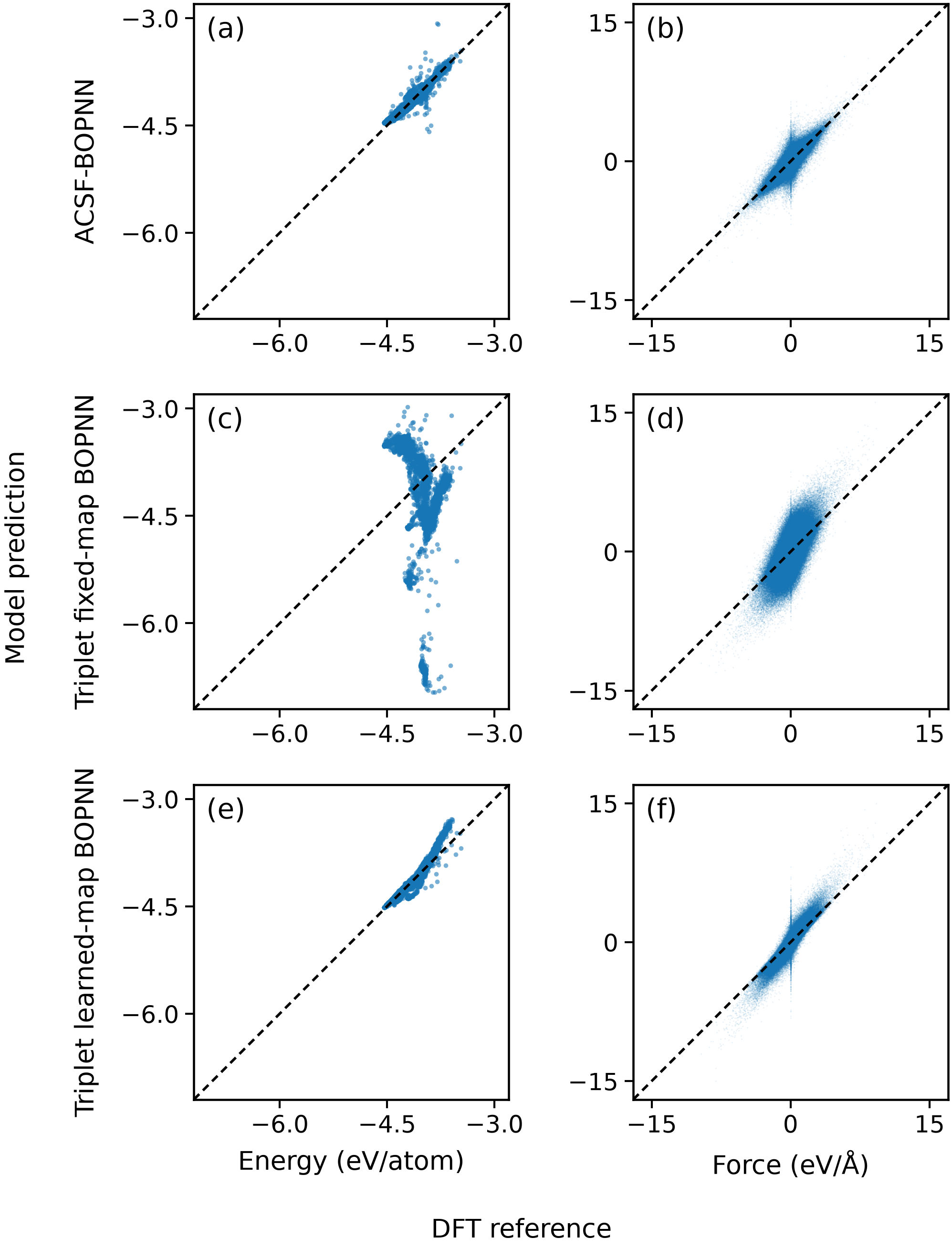}
\caption{Validation parity plots for the three BOPNN architectures. Panels (a, b), (c, d), and (e, f) show ACSF-BOPNN, triplet fixed-map BOPNN, and triplet learned-map BOPNN
, respectively; the left and right columns show energy per atom and Cartesian force components, respectively. The corresponding training-set results and stress parity plots are provided in the Supplementary Material. Dashed lines indicate exact agreement with the DFT reference values.}
\label{fig:bopnn-parity}
\end{figure}

We compared the overall fitting accuracies of the three BOPNN variants using the same validation subset before examining individual material properties. Table~\ref{tab:bopnn-rmse} summarizes the validation root-mean-square errors (RMSEs) of the three architectures. Figure ~\ref{fig:bopnn-parity} shows the corresponding parity plots. The ACSF-BOPNN and triplet learned-map BOPNN exhibited comparable overall accuracies, while the triplet fixed-map BOPNN showed substantially larger deviations. More detailed parity plots for the energy, force, and stress including the training-set results and category-resolved comparisons over the full dataset are provided in the Supplementary Material.

\begin{table}[tbp]
\caption{Validation RMSEs of the three BOPNN architectures.}
\label{tab:bopnn-rmse}
\centering
\small
\setlength{\tabcolsep}{3pt}
\resizebox{\columnwidth}{!}{%
\begin{tabular}{cccc}
\hline\hline
Model & Energy & Force & Stress \\
& (meV/atom) & (meV/\AA) & (meV/\AA$^{3}$) \\
\hline
ACSF-BOPNN & 68.6 & 455.5 & 41.9 \\
Triplet fixed-map BOPNN & 938 & 1138 & 62.6 \\
Triplet learned-map BOPNN & 83.6 & 431.1 & 8.75 \\
\hline\hline
\end{tabular}%
}
\end{table}

Despite their substantially different representations of the local atomic environment, the ACSF-BOPNN and triplet learned-map BOPNN exhibited comparable overall fitting accuracies. The ACSF-BOPNN achieved a slightly lower energy RMSE, while the triplet learned-map BOPNN showed a slightly lower force RMSE and a substantially lower stress RMSE. This result confirms that an atom-centered ACSF representation can achieve interpolation accuracy comparable to that of the bond-conditioned triplet representation for the present training dataset. By contrast, the triplet fixed-map BOPNN exhibited considerably larger errors across all three quantities. The two triplet-based models shared the same form of triplet representation and differed in mapping from aggregated effective coordination \(\zeta_{ij}\) to the bond-order factor; therefore, this result suggests that retaining analytical Tersoff mapping limits model flexibility. Meanwhile, replacing it with a learned mapping improves fitting accuracy to a level comparable to that of the ACSF-BOPNN.

\begin{figure}[tbp]

\centering
\makebox[\linewidth][l]{\textbf{(a)}}\\
\includegraphics[width=0.95\linewidth]{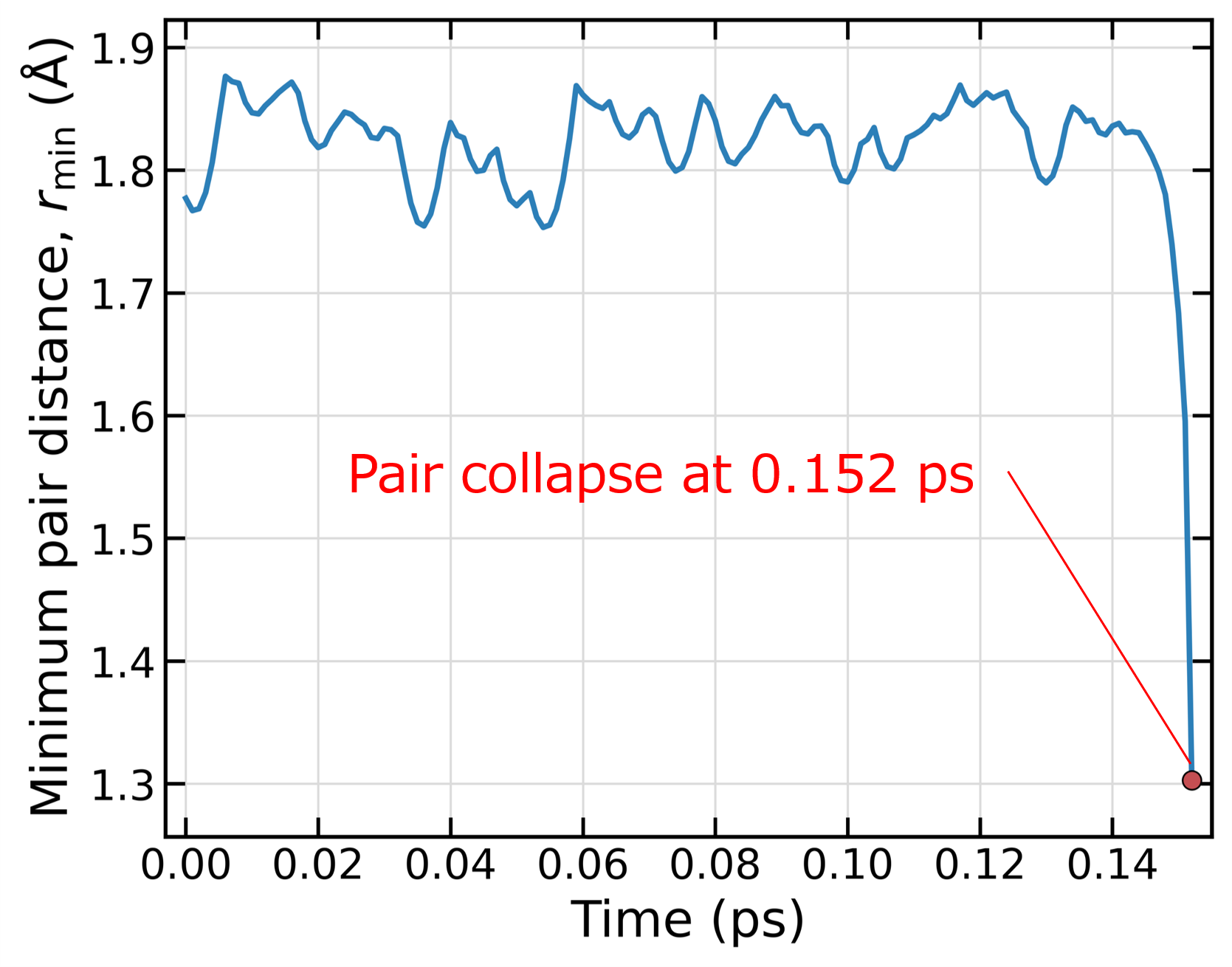}
\par\vspace{0.4em}
\makebox[\linewidth][l]{\textbf{(b)}}\\
\includegraphics[width=0.95\linewidth]{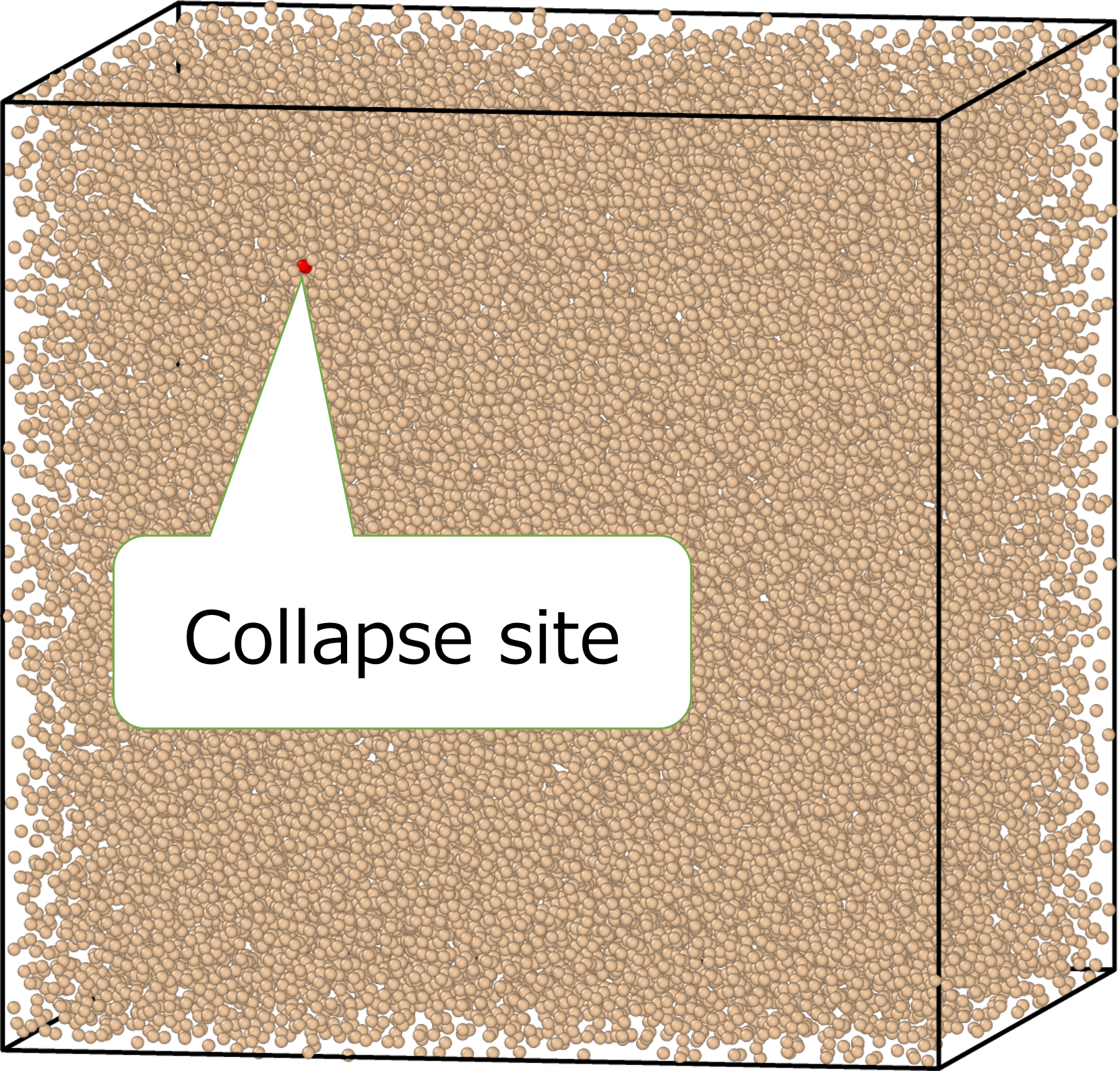}
\par\vspace{0.4em}
\makebox[\linewidth][l]{\textbf{(c)}}\\
\includegraphics[width=0.95\linewidth]{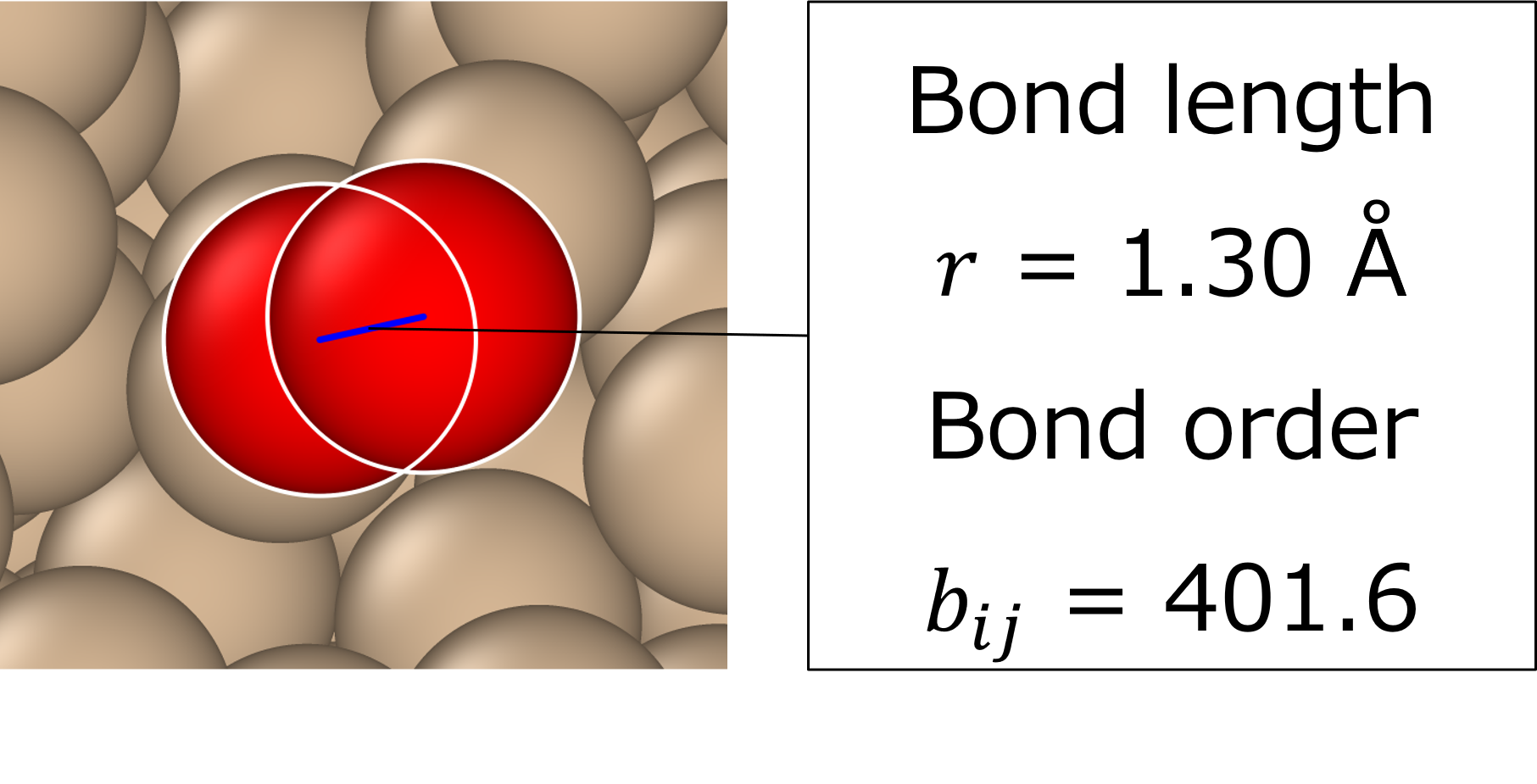}
\caption{Failure of
ACSF-BOPNN during the disordered-Si MD simulation. (a) Time evolution of the minimum interatomic distance showing an abrupt local pair collapse at 0.152~ps. (b) Atomic configuration at 0.152~ps, with the collapse site indicated. (c) Enlarged view of the collapsed pair, for which the interatomic distance decreased to 1.30~\AA\ and the predicted bond-order factor reached $b_{ij}\approx 4.0\times10^{2}$. The simulation terminated at the subsequent step (0.153~ps).}
\label{fig:nearest-bond-overlay}
\end{figure}

Despite their comparable validation errors, the ACSF-BOPNN and triplet learned-map BOPNN exhibited markedly different behaviors in NVE MD simulations starting from the same 32,768-atom disordered Si configuration. As indicated in Fig.~\ref{fig:nearest-bond-overlay}, the ACSF-BOPNN trajectory remained stable until a local atomic pair underwent an abrupt collapse at 0.152~ps. Immediately before termination, the minimum interatomic distance decreased to 1.30~\AA, while the bond-order factor predicted for the collapsed pair increased anomalously to $b_{ij}\approx 4.0\times10^{2}$. The simulation was terminated at the subsequent step (0.153~ps). By contrast, the corresponding triplet learned-map BOPNN trajectory completed the 1-ps NVE simulation without losing atoms or undergoing a comparable collapse; its minimum interatomic distance over the trajectory was 1.77~\AA.

This failure is consistent with the unconstrained bond-order prediction of the ACSF-BOPNN. Unlike the original Tersoff formulation in which \(b_{ij}\leq1\) and the two-atom case yielded \(b_{ij}=1\), the ACSF-BOPNN did not impose bounds on its output. By contrast, the triplet learned-map BOPNN constrained the bond-order response to $b_{ij}=1$ in the two-atom case and to decrease monotonically with increasing effective coordination. Therefore, this markedly different dynamical behavior suggests that comparable validation RMSEs do not necessarily imply comparable stability in MD simulations. This observation aligns with previous reports showing that low average test errors alone do not ensure reliable atomic dynamics in MD simulations\cite{liuDiscrepanciesErrorEvaluation2023}. Although this single test does not establish general robustness beyond the sampled configurations, the observed difference in the dynamical stability is consistent with the design principle of physics-informed ML potentials, wherein physically motivated model structures are used to constrain extrapolation behavior outside the training domain\cite{punPhysicallyInformedArtificial2019,kohataMachineLearnedBondOrderPotential2025}. In the triplet learned-map BOPNN, bounded and monotonic bond-order mapping excluded the observed runaway to \(b_{ij}\gg1\) by construction, providing a concrete stabilizing mechanism absent in the ACSF-BOPNN.

Considering both the fitting accuracy and dynamical stability, we adopted the triplet learned-map BOPNN as the final architecture for subsequent analyses. Hereafter, Tersoff-NN refers to the model employing this selected BOPNN architecture.

\subsection{Computational efficiency}

We benchmarked two 32,768-atom Si systems to quantify the computational performance: diamond Si at 300~K and liquid Si at 3000~K. The crystalline system provides a nearly uniform tetrahedral environment with narrow distributions of bond lengths and bond angles, while the liquid system contains fluctuating coordination numbers and a broad distribution of local geometries. Collectively, these benchmarks demonstrate the computational performance of Tersoff-NN in both uniform and strongly disordered atomic environments.

Tersoff-NN was benchmarked against classical analytical potentials and representative ML-IAP implementations; the specific potential parameterizations and model identifiers are provided in the Supplementary Material. For the primary comparison with the Behler--Parrinello-type neural-networks (BPNNs), we used separately trained in-house models implemented in SimpleNN\cite{leeSIMPLENNEfficientPackage2019} and n2p2\cite{singraberLibraryBasedLAMMPSImplementation2019}. These two packages employ different cutoff-function definitions; therefore, the same trained model cannot be transferred directly between them without changing the underlying descriptors. Both models were trained in-house using identical radial and angular symmetry-function parameters, cutoff radius, and NN architecture.

In addition, we evaluated the publicly distributed Si BPNN of Zuo \emph{et al.}\cite{zuoPerformanceCostAssessment2020} as a compact published reference. However, its trajectory exhibited a severe temperature excursion with a substantial drift in the extended-system conserved quantity under the 3000~K liquid-Si protocol. The resulting trajectory no longer represented the target thermodynamic state; therefore, its timing was excluded from the high-temperature performance comparison. The temperature and energy trajectories are reported in the Supplementary Material.

All benchmarks were performed in the CPU-only mode on a single-socket compute node equipped with an AMD EPYC 7702P processor (64 physical cores and 128 hardware threads). Each calculation used 16 MPI ranks with 1 OpenMP thread per rank, held fixed across all models. Within each state point, all models used identical simulation settings, except for the pair style and potential-specific input files. For each reported model--state-point combination, the wall-clock time reported by LAMMPS\cite{thompsonLAMMPSFlexibleSimulation2022} was measured over a 10,000-step production run. Each condition was evaluated only once; therefore, no statistical error bars are reported, and the results reflect the performance under this fixed hardware and parallel configuration, rather than individually optimized peak performance. Additional implementation details, benchmark input settings, and absolute timing data are provided in the Supplementary Material.

\begin{figure}[tbp]

\centering
\begin{minipage}[t]{0.48\linewidth}
\textbf{(a)}\\
\includegraphics[width=\linewidth]{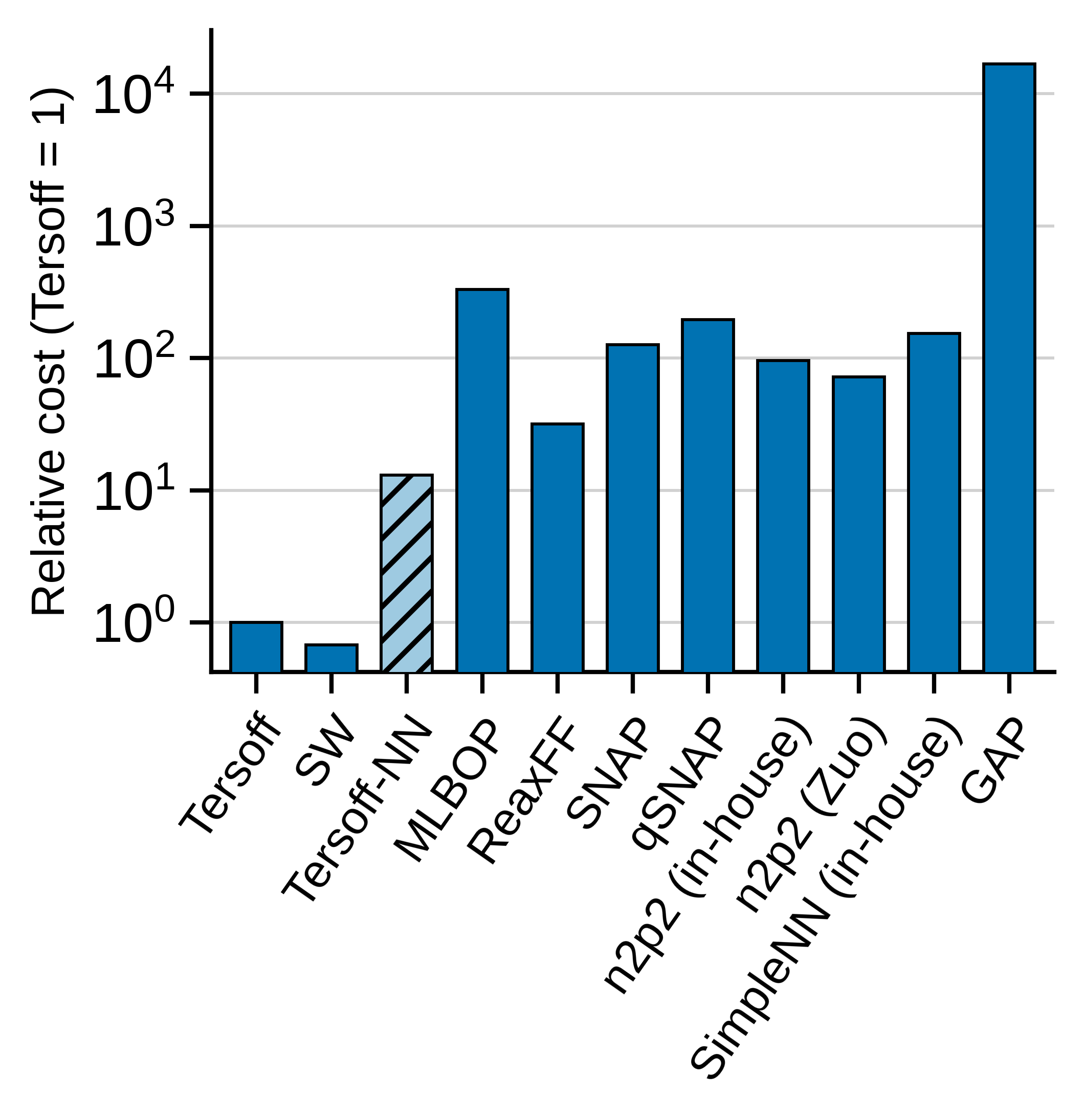}
\end{minipage}\hfill
\begin{minipage}[t]{0.48\linewidth}
\textbf{(b)}\\
\includegraphics[width=\linewidth]{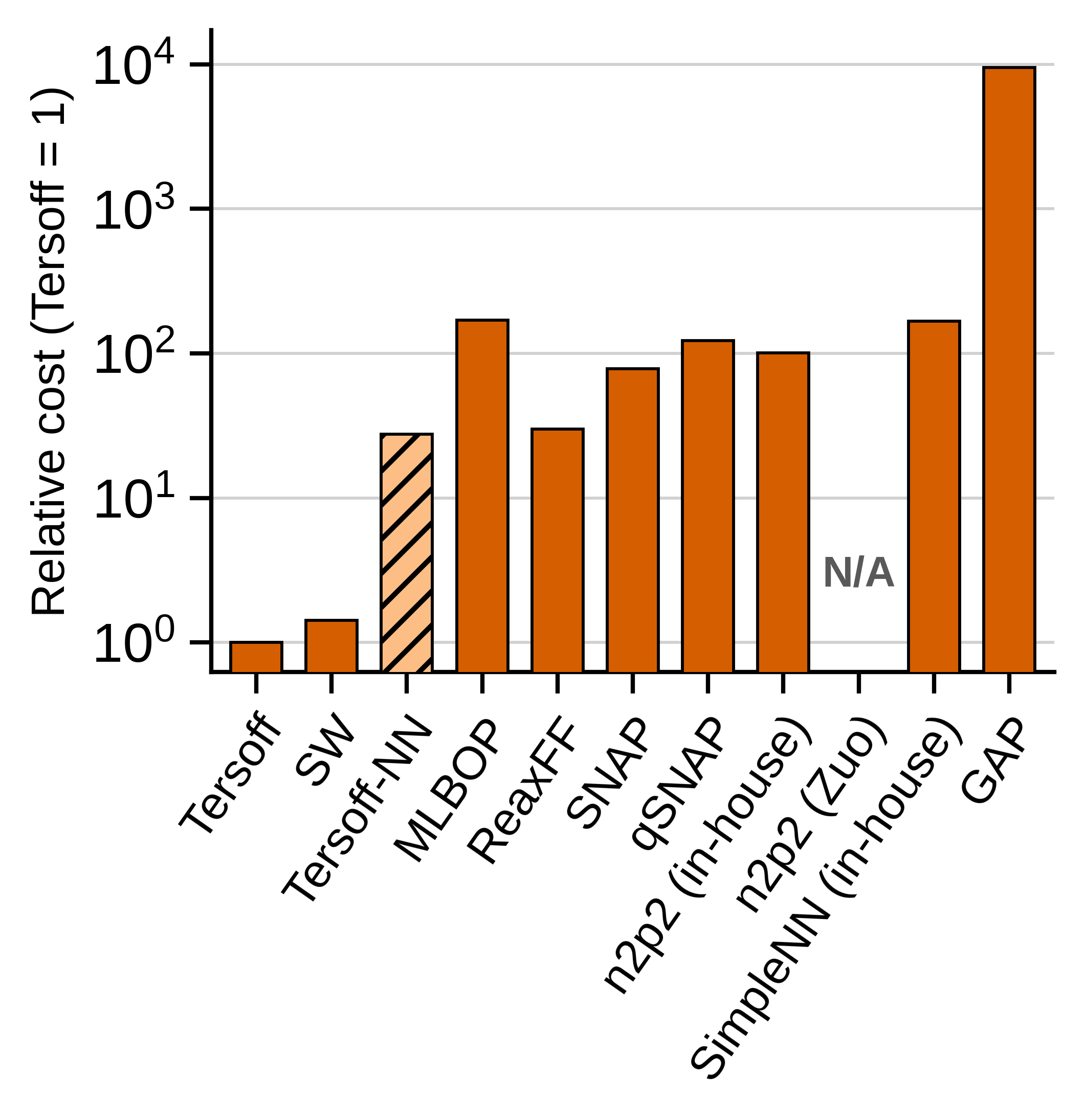}
\end{minipage}
\caption{Relative computational cost for (a) diamond Si at 300~K and (b) liquid Si at 3000~K. For each system, the wall-clock time per atom-step was normalized by the corresponding time measured with Tersoff. No value is reported for Zuo BPNN at 3000~K because its trajectory deviated from the target liquid state;
the corresponding stability diagnostics are provided in the Supplementary Material.}
\label{fig:benchmark-cost-relative}
\end{figure}

As shown in Fig.~\ref{fig:benchmark-cost-relative}, the Tersoff and Stillinger--Weber\cite{stillingerComputerSimulationLocal1985} potentials were the fastest, as expected for purely analytical IAPs. Tersoff-NN required wall-clock times of $1.43\times10^{-6}$ and $5.37\times10^{-6}$~s per atom step for diamond Si at 300~K and liquid Si at 3000~K, corresponding to throughputs of $6.97\times10^{5}$ and $1.86\times10^{5}$ atom-steps/s, respectively. The wall-clock cost was approximately 13 and 28 times that of the Tersoff potential on the diamond Si benchmark and liquid Si benchmark, respectively. Although the relative overhead was higher in the liquid environment, Tersoff-NN remained approximately 4--12 times faster than the two BPNN implementations across all tested conditions. At 300~K, Tersoff-NN was approximately \(5.5\) times faster than the compact Zuo BPNN. The Zuo model did not maintain the target 3000~K liquid state; therefore, this single-state-point result was not included in the quoted 4--12 speedup range. Details of this stability test are provided in the Supplementary Material.

The throughput benchmark compared the practical performance of the complete potential implementations; however, it did not reveal the specific computational operations driving the observed differences. Among the ML-IAPs considered, a BPNN was selected for component-level analysis because it provides the closest conventional NN counterpart to Tersoff-NN. Both models used feed-forward multilayer perceptrons to map local geometric information to energy-related quantities and obtain atomic forces by differentiating NN outputs. This common structure allows the costs of input-feature construction, NN evaluation, and force calculation to be compared using identical timing categories. The input features comprise radial and angular ACSFs in SimpleNN and bond-conditioned triplet features in Tersoff-NN.

\begin{figure}[t]
\centering
\includegraphics[width=\linewidth]{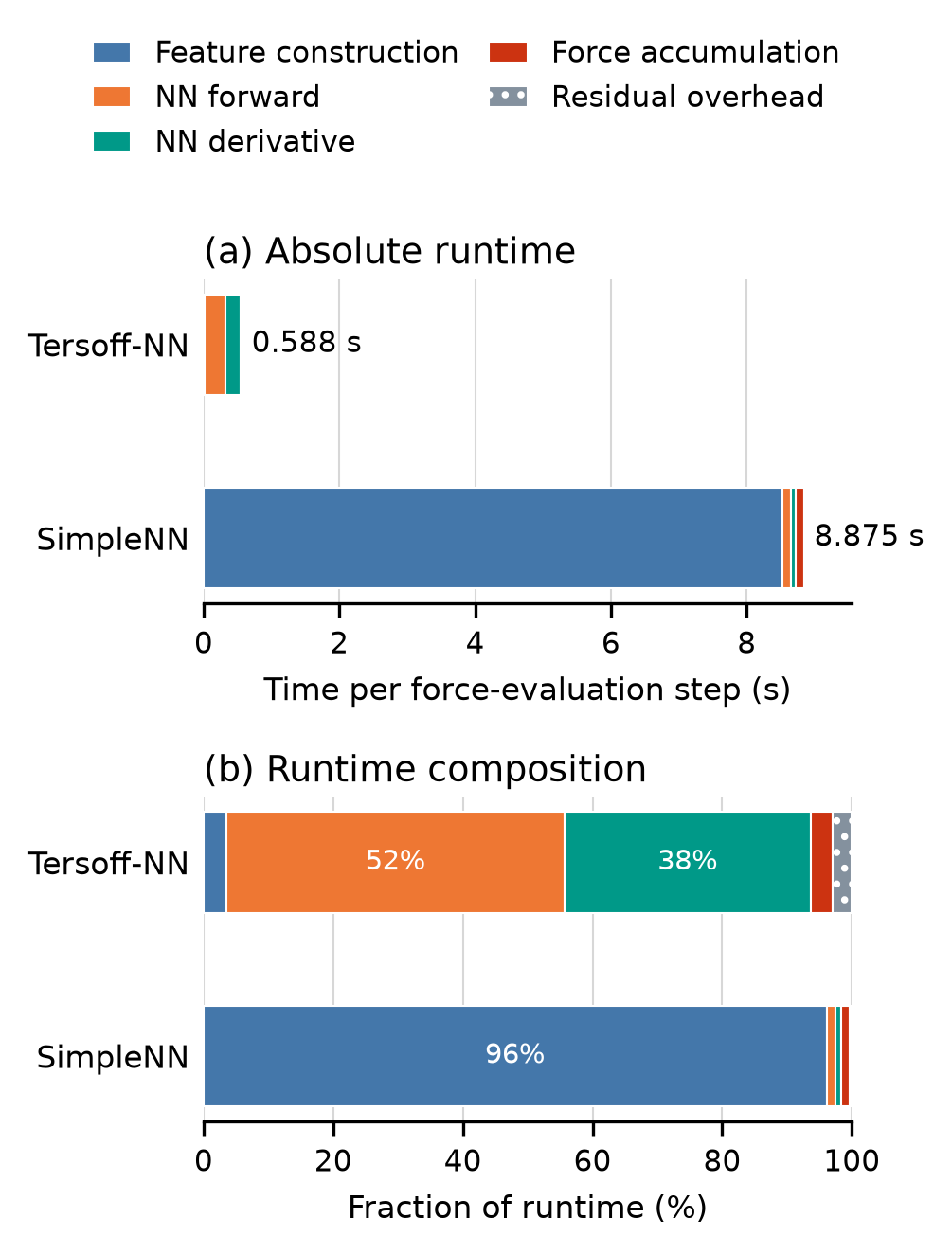}
\caption{Internal timing decomposition for the BPNN implemented in SimpleNN and the direct Tersoff-NN implementation. (a) Absolute time per force-evaluation step and (b)
the corresponding fractional composition. Reported totals are shown beside bars; component means sum to totals within rounding. Values represent means over 100 repeated evaluations with fixed atomic positions, excluding the initialization evaluation at step 0. Standard deviations are smaller than 0.006~s for all entries.}
\label{fig:timing-decomposition}
\end{figure}

Accordingly, we instrumented the Tersoff-NN and SimpleNN pair styles and separated the total force-evaluation time into input-feature construction, NN forward evaluation, NN derivatives, force accumulation, and residual overhead. A fixed liquid configuration was evaluated for 100 steps without time integration, and the averages are reported in Fig.~\ref{fig:timing-decomposition}. SimpleNN was used for this analysis because its pair style could be instrumented at the required level without modifying the optimized n2p2 implementation. Therefore, the decomposition should be interpreted as an implementation-level bottleneck analysis, rather than a general performance ranking of ML-IAP formalisms.

The two models exhibited qualitatively different computational bottlenecks. In SimpleNN, input-feature construction required 8.533~s of the total 8.875~s per step (approximately 96\% of the runtime), while NN forward and derivative evaluations required only 0.193~s. This dominant feature-construction cost included those for evaluation of the radial and angular ACSFs and construction of their coordinate-derivative buffers for force calculation. This result is consistent with previous analyses of artificial NN IAPs identifying structural-fingerprint evaluation, rather than NN propagation, as the primary computational cost for typical network architectures\cite{dickelLAMMPSImplementationRapid2021}.

In Tersoff-NN, input-feature construction required only 0.021~s per step. Most of the total runtime was spent on NN forward evaluation (0.307~s) and derivative evaluation (0.224~s). Therefore, NN-related calculations are not intrinsically cheaper in Tersoff-NN than those in the SimpleNN benchmark. In the instrumented SimpleNN comparison, the lower total cost was attributed to avoiding repeated construction and differentiation of radial and angular ACSFs. Further, the production benchmark against the optimized n2p2 demonstrated that the overall speed advantage was not limited to SimpleNN; however, its internal origin was not resolved separately. These results show that the present hybrid Tersoff-NN implementation retains a substantial computational advantage over the two in-house BPNN models under the tested conditions.

\subsection{Crystalline properties}

\begin{table*}[tbp]
\caption{Equilibrium structural parameters, bulk moduli, and elastic constants of diamond Si.}
\begin{ruledtabular}
\begin{tabular}{cccccccc}
Method & $a_{0}$ (\AA) & $V_{0}$ (\AA$^{3}$/atom) & $B_{0}^{\mathrm{BM}}$ (GPa) & $C_{11}$ (GPa) & $C_{12}$ (GPa) & $C_{44}$ (GPa) & $B^{\mathrm{elastic}}$ (GPa) \\
DFT & 5.473 & 20.497 & 88.0 & 154.0 & 57.6 & 99.8 & 89.7 \\
Tersoff & 5.432 & 20.036 & 97.6 & 135.5 & 67.8 & 112.9 & 90.4 \\
Pre-NN Tersoff & 5.468 & 20.440 & 80.4 & 106.8 & 68.4 & 91.5 & 81.2 \\
Tersoff-NN & 5.464 & 20.388 & 74.5 & 106.1 & 58.8 & 88.0 & 74.6 \\
\end{tabular}
\end{ruledtabular}
\label{tab:diamond-properties}
\end{table*}

\begin{figure}[tbp]
\centering
\includegraphics[width=\linewidth]{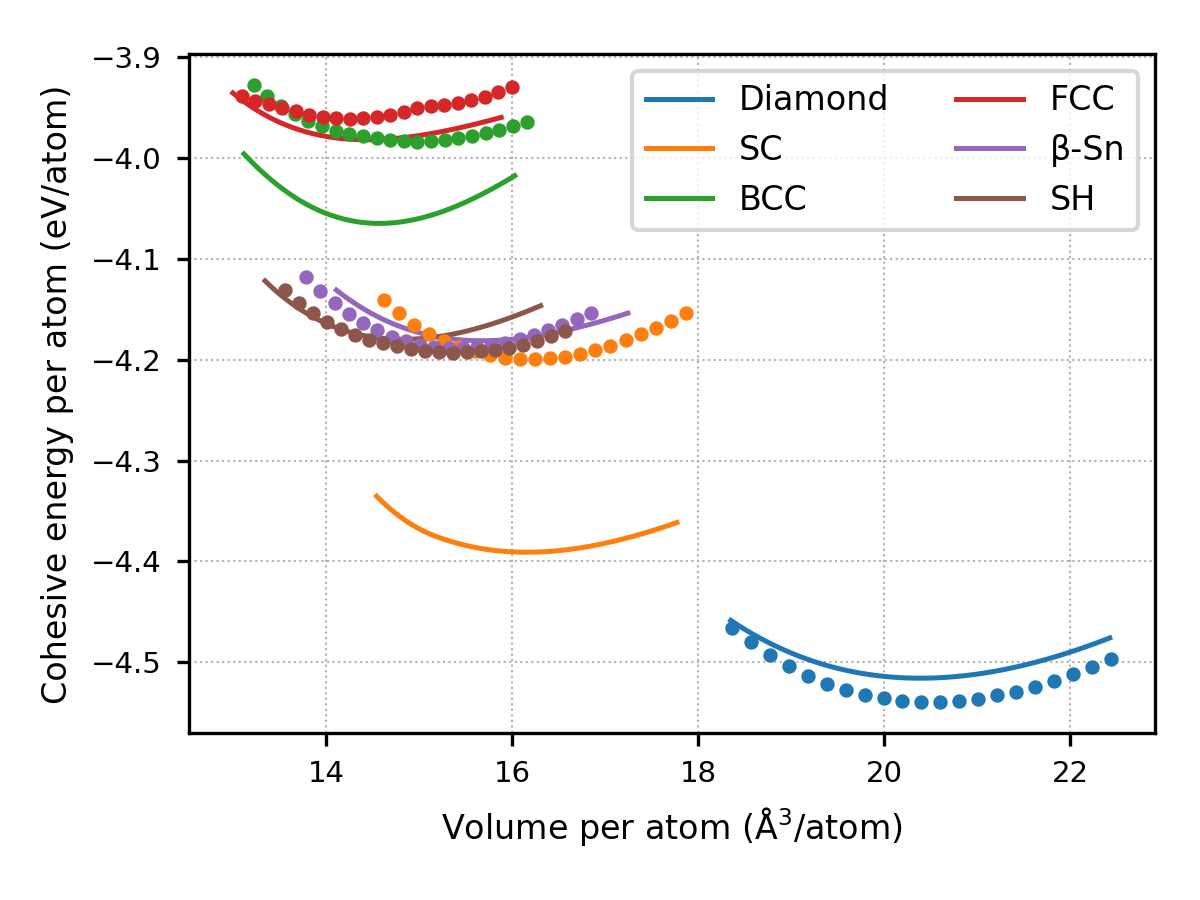}
\caption{Cohesive-energy--volume relationships for diamond, simple-cubic (SC), body-centered-cubic (BCC), face-centered-cubic (FCC), $\beta$-Sn, and simple-hexagonal (SH) Si. Symbols denote DFT reference values, and solid curves represent the corresponding Tersoff-NN results obtained by first optimizing only the reference volume isotropically at zero pressure with the atoms fixed, followed by evaluation of single-point energies for isotropically scaled cells at fixed cell shape and fractional coordinates.}
\label{fig:ev-curves}
\end{figure}

Fig.~\ref{fig:ev-curves} compares the energy--volume relationships of several Si polymorphs calculated using DFT and Tersoff-NN. Tersoff-NN reproduced the equilibrium volume of diamond Si satisfactorily; however, its fitted bulk modulus was lower than the DFT value. Furthermore, reasonable agreement was obtained near the energy minima of the $\beta$-Sn and simple-hexagonal (SH) structures, both of which were experimentally observed high-pressure phases of Si\cite{huPhasesSiliconHigh1984,mcmahonNewHighpressurePhase1993}. By contrast, larger deviations were observed for the simple-cubic (SC), body-centered-cubic (BCC), face-centered-cubic (FCC) structures, for which Tersoff-NN predicted substantially lower energies than those obtained using DFT. However, the qualitative energy ordering of these structures was retained.

The energy--volume curves were fitted using the third-order Birch--Murnaghan equation of state\cite{birchFiniteElasticStrain1947} to quantify the agreement near the equilibrium diamond structure. For this comparison, we used the model obtained after preliminary optimization of the analytical parameters but before NN training, hereafter referred to as ``Pre-NN Tersoff.'' This model incorporated the optimized radial parameters and extended cutoff used in Tersoff-NN while retaining the original Tersoff bond-order expression. The fitting procedure and optimized parameters are provided in the Supplementary Material. The resulting equilibrium structural parameters and bulk moduli are summarized in Table~\ref{tab:diamond-properties}.

The equilibrium geometry of diamond Si was well reproduced by both Pre-NN Tersoff and Tersoff-NN. However, the corresponding bulk moduli and elastic constants exhibited larger deviations from the respective DFT values. Preliminary optimization of radial parameters substantially softened the diamond response. Moreover, subsequent NN training improved $C_{12}$ but did not recover the reduction in $C_{11}$. This behavior is consistent with the fitting procedure used for the analytical parameters: these parameters were optimized against the cohesive-energy curve over a wide range of isotropic strains without explicitly constraining the local second derivatives at equilibrium. Consequently, reproducing the global energy--volume relationship of diamond Si over a broad strain range does not necessarily ensure an accurate description of the local elastic curvature near equilibrium.

\subsection{Finite-temperature dynamics and disordered Si}

We examined the finite-temperature vibrational behavior of crystalline Si before evaluating disordered phases. The spectral energy density (SED) spectra calculated from the finite-temperature MD trajectories\cite{thomasPredictingPhononDispersion2010} indicate that the spectral peaks shift toward lower frequencies (phonon softening) and broaden with increasing temperature. The corresponding SED maps are provided in the Supplementary Material.

\begin{figure}[tbp]

\centering
\makebox[\linewidth][l]{\textbf{(a)}}\\
\includegraphics[width=\linewidth]{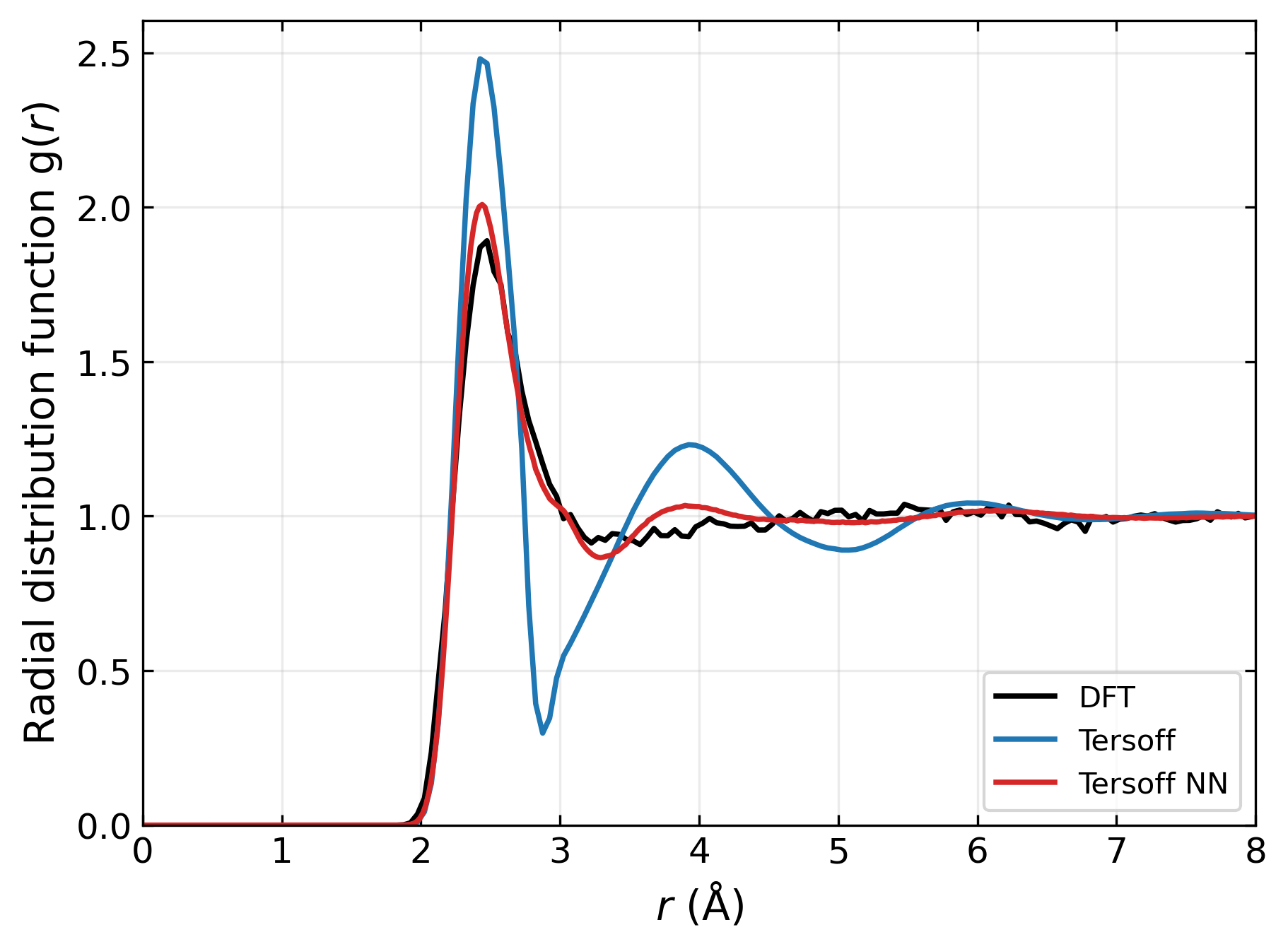}
\par\vspace{0.3em}
\makebox[\linewidth][l]{\textbf{(b)}}\\
\includegraphics[width=\linewidth]{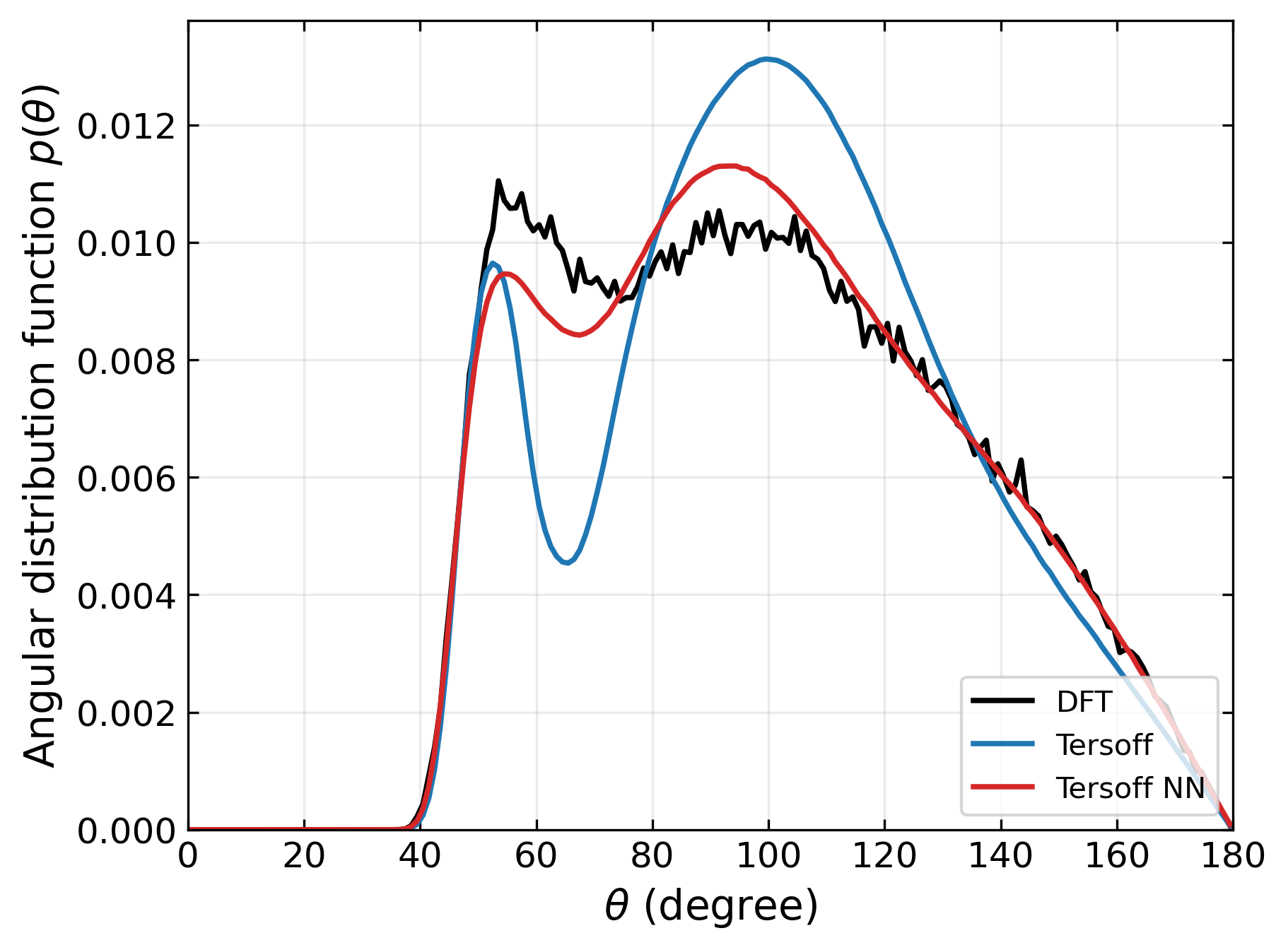}
\caption{Structural distributions of liquid Si at 3000~K obtained from DFT, Tersoff
, and Tersoff-NN: (a) radial distribution function (RDF) and (b) angular distribution function (ADF).}
\label{fig:liquid-rdf-adf}
\end{figure}

Subsequently, we focused on liquid Si, which provides a more direct test of the learned bond-order response under strongly distorted and rapidly fluctuating local environments. Fig.~\ref{fig:liquid-rdf-adf} compares the radial and angular distribution functions obtained from DFT, the Tersoff potential, and Tersoff-NN at 3000~K. As shown in Fig.~\ref{fig:liquid-rdf-adf}(a), the Tersoff potential produced an excessively sharp first radial distribution function (RDF) peak and a deep first minimum, indicative of an overstructured liquid network. Meanwhile, Tersoff-NN suppressed this artificial short-range order and yielded an RDF close to the DFT reference. The first minimum of the Tersoff-NN RDF appeared at approximately $3.29$~\AA{}; direct neighbor counting up to $3.2$~\AA{} yielded an average coordination number of 6.55, close to the DFT value of 6.37. Comparison of the angular distribution function (ADF) in Fig.~\ref{fig:liquid-rdf-adf}(b) revealed that Tersoff-NN exhibited two broad peaks at approximately $54.5^\circ$ and $93.5^\circ$. This two-peak feature is consistent with the ab initio MD (AIMD) results of liquid Si reported by Stich, Car, and Parrinello, wherein they observed broad angular maxima around $60^\circ$ and $90^\circ$ with a broad coordination distribution centered near the sixfold coordination \cite{stichStructuralBondingDynamical1991}. The results of the further analyses of the coordination-number and coordination-resolved angular distributions are provided in the Supplementary Material.

\begin{figure}[tbp]
\centering
\makebox[\linewidth][l]{\textbf{(a)}}\\
\includegraphics[width=\linewidth]{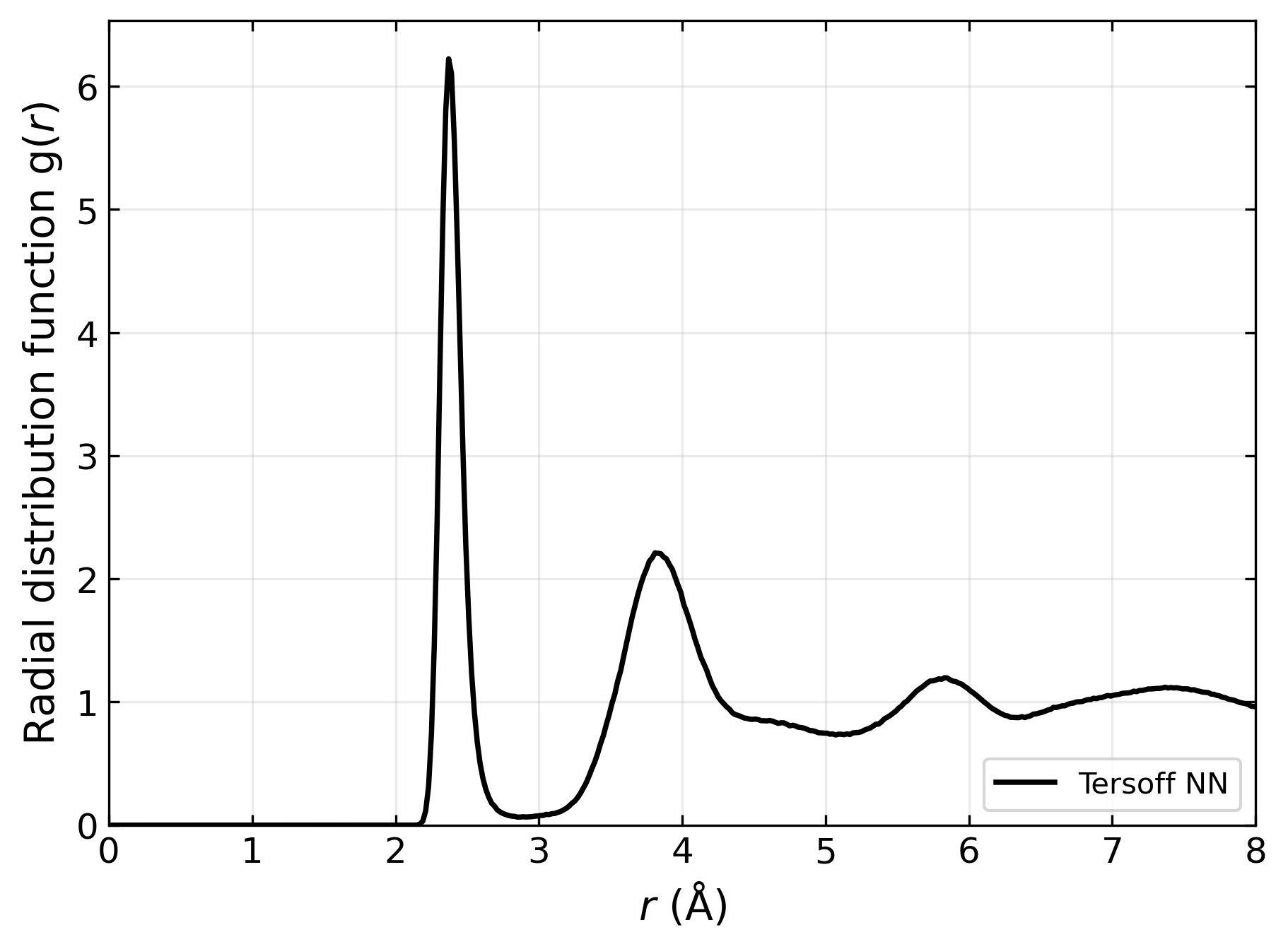}
\par\vspace{0.3em}
\makebox[\linewidth][l]{\textbf{(b)}}\\
\includegraphics[width=\linewidth]{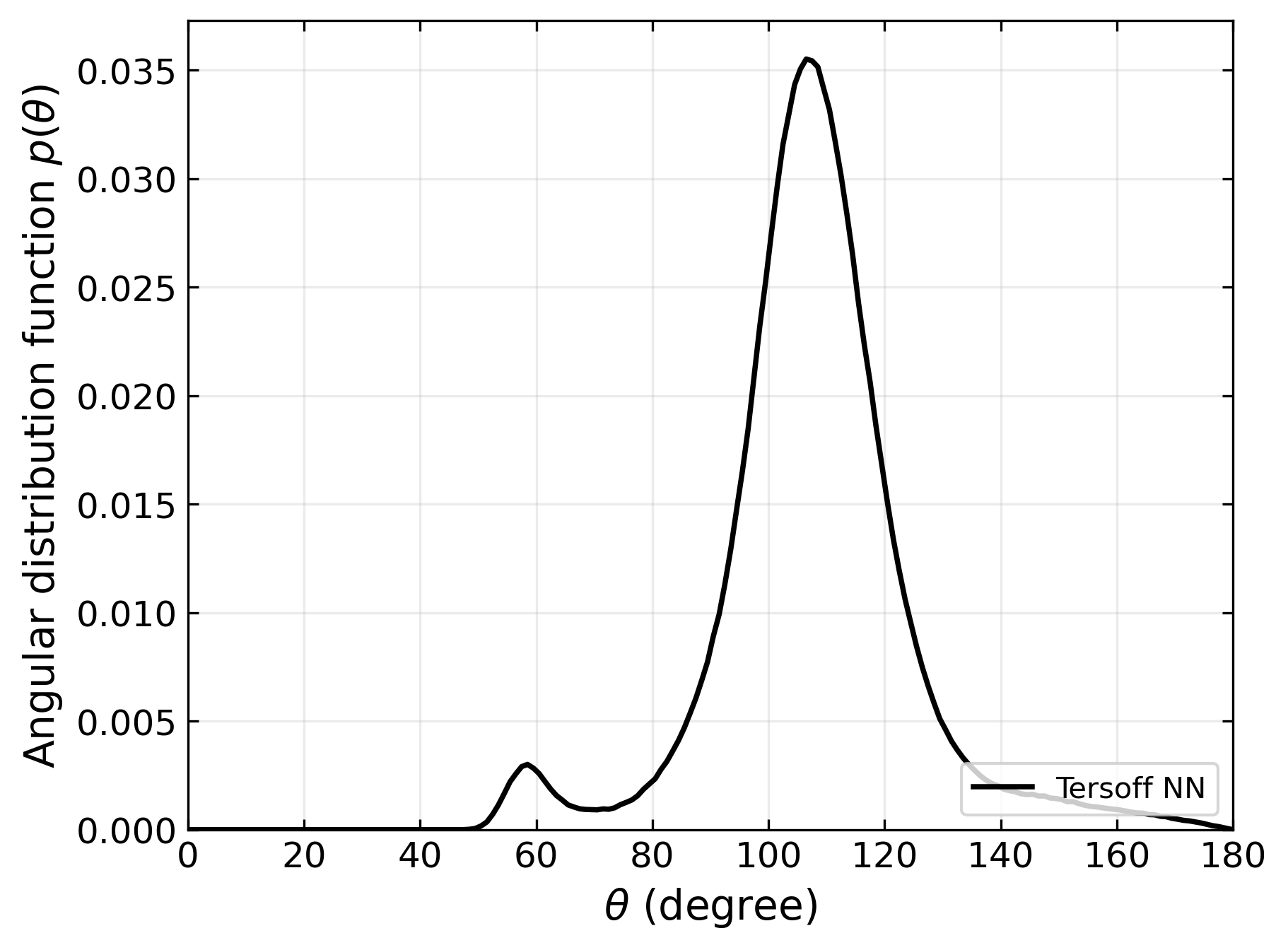}
\caption{Structural distributions of the final melt--quenched disordered Si structure: (a)
RDF and (b) ADF. ADF was evaluated using a neighbor cutoff of 2.85~\AA.}
\label{fig:mq-rdf-adf}
\end{figure}

We performed an NPT melt--quench simulation using a 4096-atom Si cell to examine whether Tersoff-NN could generate and stabilize a disordered solid structure after cooling from the liquid state. The complete heating and cooling protocols as well as the resulting potential-energy hysteresis are presented in the Supplementary Material. The resulting quenched structure had a density of 2.32~g\,cm$^{-3}$, slightly higher than the experimental value of approximately 2.29~g\,cm$^{-3}$ reported for ion-implanted amorphous Si\cite{custerDensityAmorphousSi1994}. Using a first-neighbor cutoff of 2.85~\AA{} employed in previous studies of amorphous Si\cite{santosGenerationAmorphousSi2019,deringerRealisticAtomisticStructure2018}, the average coordination number of the final structure was calculated to be 4.23. Thus, the quenched structure is slightly denser than the experimentally characterized amorphous Si and exhibits an average coordination number higher than the ideal fourfold value.

Fig.~\ref{fig:mq-rdf-adf} shows the RDF and ADF of the final melt--quenched structure. The RDF exhibited a sharp first-neighbor peak near 2.37~\AA\ and a first minimum at approximately 2.89~\AA, followed by damped oscillations approaching unity at longer distances. This behavior confirms the loss of the long-range crystalline order while retaining pronounced short-range correlations.

The ADF was evaluated using a neighbor cutoff of 2.85~\AA{}, consistent with the coordination analysis. The distribution was dominated by a broad peak near the tetrahedral angle and a weaker secondary peak at approximately $60^\circ$. The primary peak indicates the presence of tetrahedral-like local environments in the quenched structure, while the secondary peak suggests the coexistence of compact, overcoordinated motifs. Although the melt--quench simulation yields a structurally disordered solid, the resulting network is not a fully developed tetrahedral amorphous Si structure.

\subsection{Exploratory out-of-domain assessment on reconstructed Si surfaces}

We examined the reconstructed Si surfaces excluded from training as an exploratory out-of-domain benchmark. Although several characteristic motifs were retained after static relaxation, substantial quantitative deviations and finite-temperature instabilities were observed.

Tests on several simpler reconstructed surfaces revealed a consistent trend: Tersoff-NN retained the characteristic surface-bonding motifs of the reconstructed surfaces while failing to quantitatively reproduce some reconstruction-specific atomic relaxations. For example, the Si (100) dimers remained symmetric rather than buckled\cite{wolkowDirectObservationIncrease1992}; vertical displacement within the Pandey chain was substantially smaller than the experimental and first-principles values\cite{xuAtomicStructureCleaved2004,zitzlspergerInitioCalculationStructure1997a}; and downward relaxation beneath Si (111) adatoms was smaller than the previous first-principles values\cite{meadeAdatomsSi111Ge1111989,solaresDensityFunctionalTheory2005}. Furthermore, short NVT simulations at 300~K revealed that the Si (100) dimers, Pandey chains, and most T$_4$-like adatom motifs remained stable over 10~ps; however, a partial rearrangement was observed for the $c(2\times8)$ structure. Relaxed structures and finite-temperature structural changes are presented in the Supplementary Material.

\begin{figure}[tbp]
  \centering
  \includegraphics[width=\linewidth]{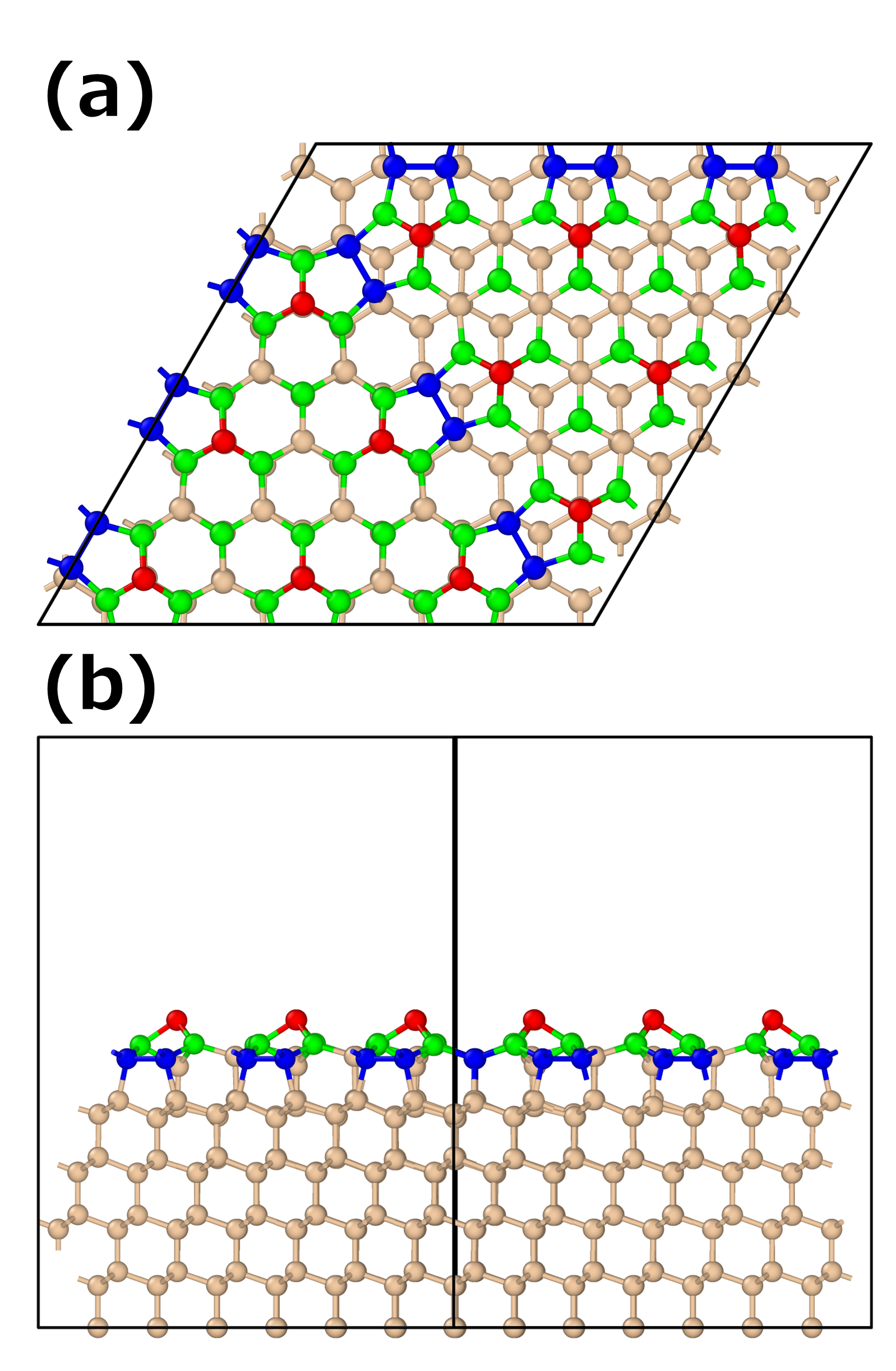}
  \caption{Si (111)-$(7\times7)$ dimer--adatom--stacking-fault (DAS) structure after static relaxation with Tersoff-NN: (a) top view and (b) side view. Adatoms, surface Si atoms bonded to adatoms, and dimer atoms are shown in red, green, and blue, respectively, and the remaining Si atoms are shown in beige. For visual clarity, only the upper six Si layers are shown in (a), while all 12 Si layers included in the calculation are shown in (b).}
  \label{fig:das_optimize}
  \end{figure}

The complex Si (111)-$(7\times7)$ dimer--adatom--stacking-fault (DAS) reconstruction\cite{takayanagiStructureAnalysisSi11171985} provides a stricter out-of-domain test. Starting from the preconstructed DAS model, static relaxation with Tersoff-NN preserved all 12 adatoms and 9 dimers. The same topology was retained by the conventional Tersoff potential, consistent with earlier Tersoff-based studies of the Si (111)-$(7\times7)$ DAS structure\cite{liEnergyVibrationalSpectrum1988a}. This indicates that preservation of the qualitative DAS framework is not unique to the NN model. However, Tersoff-NN reproduced the PBE-relaxed geometry more accurately: across the top five layers, the atomic-position root-mean-square deviation decreased from 0.151~\AA\ for the Tersoff potential to 0.103~\AA\ for Tersoff-NN. Additionally, significant improvements were obtained for the adatom and dimer environments, while the largest positional errors were observed for the rest atoms. A detailed structural comparison, including the earlier Tersoff parameterization used by Li et al., is provided in the Supplementary Material.

Overall, these surface tests revealed that Tersoff-NN successfully retained characteristic motifs in low-coordination environments outside its training domain. The characteristic reconstruction motifs were retained after static relaxation, and simpler reconstructed surfaces remained largely stable during short simulations at 300~K. Although the more complex Si (111)-$(7\times7)$ DAS structure was preserved after static relaxation, multiple adatoms were displaced from their initial sites during the 300~K simulation. These results indicate that the Tersoff-type framework provides a useful structural basis to describe previously unseen surface environments, thereby enabling the characteristic reconstruction motifs to be retained even without surface configurations in the training dataset. However, the finite-temperature instability of the Si (111)-$(7\times7)$ DAS adatom arrangement indicates that additional surface configurations are required for a quantitatively reliable description of more complex reconstructed surfaces and their dynamics.

\section{Conclusions}
In this study, we developed Tersoff-NN---a hybrid IAP for Si, which retains the analytical radial terms and cutoff structure of the Tersoff potential while learning only the bond-order factor using compact NNs. This constrained bond-order architecture achieved a fitting accuracy comparable to that of the ACSF-based model while facilitating a substantially more stable MD behavior, demonstrating that physically motivated constraints could improve dynamic stability in configurations not adequately characterized by validation errors alone. Furthermore, Tersoff-NN remained approximately 4--12 times faster than the BPNN implementations by avoiding construction and differentiation of atom-centered symmetry functions.

Despite its compact architecture, Tersoff-NN reproduced the equilibrium geometry of diamond Si and improved the short-range structure of liquid Si relative to the conventional Tersoff potential. Melt--quench simulations yielded a disordered solid with predominantly four-coordinated, tetrahedral-like local environments, while the reconstructed surfaces excluded from training retained the characteristic structural motifs after relaxation. However, some surface structures exhibited quantitative deviations and finite-temperature instabilities, indicating that additional surface-specific training data are required for reliable surface simulations. Overall, these findings demonstrate that restricting ML to the bond-order term offers a practical strategy for improving analytical IAPs while maintaining physical constraints and computational efficiency.

\section*{Supplementary Material}
Additional computational details and supporting analyses are provided in the Supplementary Material, available as the ancillary file \texttt{tersoffnn\_supplementary\_v3.pdf}.

\section*{Acknowledgments}

This work was supported by JSPS KAKENHI Grant no. 25K01295 and JST SPRING Grant no. JPMJSP2128.

\section*{Author Declarations}

\subsection*{Conflict of Interest}

The authors have no conflicts to disclose.

\subsection*{Author Contributions}
Yusuke Nishimura: Formal analysis, Investigation, Methodology, Software, Validation, Visualization, Writing -- original draft, Writing -- review \& editing. Sho Kubota: Investigation, Writing -- review \& editing. Kotaro Takematsu: Software, Writing -- review \& editing. Kentaro Hirai: Methodology, Software, Writing -- review \& editing. Takanobu Watanabe: Conceptualization, Funding acquisition, Resources, Supervision, Writing -- review \& editing.

\section*{Data Availability}

The data, source code, and trained models supporting the findings of this study are openly available in Zenodo at https://doi.org/10.5281/zenodo.23290766~\cite{nishimuraTersoffNNZenodo2026}.
Further data are available from the corresponding author (Y.N.) on request.

\bibliography{reference_editage_v2_au}

%aipnum4-2.bst 2019-01-14 (MD) hand-edited version of apsrev4-1.bst
\providecommand{\authorjournal}[2]{#1}
\begin{thebibliography}{54}%
\makeatletter
\providecommand \@ifxundefined [1]{%
 \@ifx{#1\undefined}
}%
\providecommand \@ifnum [1]{%
 \ifnum #1\expandafter \@firstoftwo
 \else \expandafter \@secondoftwo
 \fi
}%
\providecommand \@ifx [1]{%
 \ifx #1\expandafter \@firstoftwo
 \else \expandafter \@secondoftwo
 \fi
}%
\providecommand \natexlab [1]{#1}%
\providecommand \enquote  [1]{``#1''}%
\providecommand \bibnamefont  [1]{#1}%
\providecommand \bibfnamefont [1]{#1}%
\providecommand \citenamefont [1]{#1}%
\providecommand \href@noop [0]{\@secondoftwo}%
\providecommand \href [0]{\begingroup \@sanitize@url \@href}%
\providecommand \@href[1]{\@@startlink{#1}\@@href}%
\providecommand \@@href[1]{\endgroup#1\@@endlink}%
\providecommand \@sanitize@url [0]{\catcode `\\12\catcode `\$12\catcode
  `\&12\catcode `\#12\catcode `\^12\catcode `\_12\catcode `\%12\relax}%
\providecommand \@@startlink[1]{}%
\providecommand \@@endlink[0]{}%
\providecommand \url  [0]{\begingroup\@sanitize@url \@url }%
\providecommand \@url [1]{\endgroup\@href {#1}{\urlprefix }}%
\providecommand \urlprefix  [0]{URL }%
\providecommand \Eprint [0]{\href }%
\providecommand \doibase [0]{https://doi.org/}%
\providecommand \selectlanguage [0]{\@gobble}%
\providecommand \bibinfo  [0]{\@secondoftwo}%
\providecommand \bibfield  [0]{\@secondoftwo}%
\providecommand \translation [1]{[#1]}%
\providecommand \BibitemOpen [0]{}%
\providecommand \bibitemStop [0]{}%
\providecommand \bibitemNoStop [0]{.\EOS\space}%
\providecommand \EOS [0]{\spacefactor3000\relax}%
\providecommand \BibitemShut  [1]{\csname bibitem#1\endcsname}%
\let\auto@bib@innerbib\@empty
%</preamble>
\bibitem [{\citenamefont {Zuo}\ \emph {et~al.}(2020)\citenamefont {Zuo},
  \citenamefont {Chen}, \citenamefont {Li}, \citenamefont {Deng}, \citenamefont
  {Chen}, \citenamefont {Behler}, \citenamefont {Cs{\'a}nyi}, \citenamefont
  {Shapeev}, \citenamefont {Thompson}, \citenamefont {Wood},\ and\
  \citenamefont {Ong}}]{zuoPerformanceCostAssessment2020}%
  \BibitemOpen
  \bibfield  {author} {\bibinfo {author} {\bibfnamefont {Y.}~\bibnamefont
  {Zuo}}, \bibinfo {author} {\bibfnamefont {C.}~\bibnamefont {Chen}}, \bibinfo
  {author} {\bibfnamefont {X.}~\bibnamefont {Li}}, \bibinfo {author}
  {\bibfnamefont {Z.}~\bibnamefont {Deng}}, \bibinfo {author} {\bibfnamefont
  {Y.}~\bibnamefont {Chen}}, \bibinfo {author} {\bibfnamefont {J.}~\bibnamefont
  {Behler}}, \bibinfo {author} {\bibfnamefont {G.}~\bibnamefont {Cs{\'a}nyi}},
  \bibinfo {author} {\bibfnamefont {A.~V.}\ \bibnamefont {Shapeev}}, \bibinfo
  {author} {\bibfnamefont {A.~P.}\ \bibnamefont {Thompson}}, \bibinfo {author}
  {\bibfnamefont {M.~A.}\ \bibnamefont {Wood}},\ and\ \bibinfo {author}
  {\bibfnamefont {S.~P.}\ \bibnamefont {Ong}},\ }\bibfield  {title} {\enquote
  {\bibinfo {title} {Performance and {{Cost Assessment}} of {{Machine Learning
  Interatomic Potentials}}},}\ }\href
  {https://doi.org/10.1021/acs.jpca.9b08723} {\bibfield  {journal} {\bibinfo
  {journal} {\authorjournal{J. Phys. Chem. A}{The Journal of Physical Chemistry
  A}}\ }\textbf {\bibinfo {volume} {124}},\ \bibinfo {pages} {731--745}
  (\bibinfo {year} {2020})}\BibitemShut {NoStop}%
\bibitem [{\citenamefont {Behler}\ and\ \citenamefont
  {Parrinello}(2007)}]{behlerGeneralizedNeuralNetworkRepresentation2007}%
  \BibitemOpen
  \bibfield  {author} {\bibinfo {author} {\bibfnamefont {J.}~\bibnamefont
  {Behler}}\ and\ \bibinfo {author} {\bibfnamefont {M.}~\bibnamefont
  {Parrinello}},\ }\bibfield  {title} {\enquote {\bibinfo {title} {Generalized
  {{Neural-Network Representation}} of {{High-Dimensional Potential-Energy
  Surfaces}}},}\ }\href {https://doi.org/10.1103/PhysRevLett.98.146401}
  {\bibfield  {journal} {\bibinfo  {journal} {\authorjournal{Phys. Rev.
  Lett.}{Physical Review Letters}}\ }\textbf {\bibinfo {volume} {98}},\
  \bibinfo {pages} {146401} (\bibinfo {year} {2007})}\BibitemShut {NoStop}%
\bibitem [{\citenamefont
  {Behler}(2011)}]{behlerAtomcenteredSymmetryFunctions2011}%
  \BibitemOpen
  \bibfield  {author} {\bibinfo {author} {\bibfnamefont {J.}~\bibnamefont
  {Behler}},\ }\bibfield  {title} {\enquote {\bibinfo {title} {Atom-centered
  symmetry functions for constructing high-dimensional neural network
  potentials},}\ }\href {https://doi.org/10.1063/1.3553717} {\bibfield
  {journal} {\bibinfo  {journal} {\authorjournal{J. Chem. Phys.}{The Journal of
  Chemical Physics}}\ }\textbf {\bibinfo {volume} {134}},\ \bibinfo {pages}
  {074106} (\bibinfo {year} {2011})}\BibitemShut {NoStop}%
\bibitem [{\citenamefont {Singraber}, \citenamefont {Behler},\ and\
  \citenamefont
  {Dellago}(2019)}]{singraberLibraryBasedLAMMPSImplementation2019}%
  \BibitemOpen
  \bibfield  {author} {\bibinfo {author} {\bibfnamefont {A.}~\bibnamefont
  {Singraber}}, \bibinfo {author} {\bibfnamefont {J.}~\bibnamefont {Behler}},\
  and\ \bibinfo {author} {\bibfnamefont {C.}~\bibnamefont {Dellago}},\
  }\bibfield  {title} {\enquote {\bibinfo {title} {Library-{{Based}}
  {{{\emph{LAMMPS}}}} {{Implementation}} of {{High-Dimensional Neural Network
  Potentials}}},}\ }\href {https://doi.org/10.1021/acs.jctc.8b00770} {\bibfield
   {journal} {\bibinfo  {journal} {\authorjournal{J. Chem. Theory
  Comput.}{Journal of Chemical Theory and Computation}}\ }\textbf {\bibinfo
  {volume} {15}},\ \bibinfo {pages} {1827--1840} (\bibinfo {year}
  {2019})}\BibitemShut {NoStop}%
\bibitem [{\citenamefont {Bart{\'o}k}\ \emph {et~al.}(2010)\citenamefont
  {Bart{\'o}k}, \citenamefont {Payne}, \citenamefont {Kondor},\ and\
  \citenamefont {Cs{\'a}nyi}}]{bartokGaussianApproximationPotentials2010}%
  \BibitemOpen
  \bibfield  {author} {\bibinfo {author} {\bibfnamefont {A.~P.}\ \bibnamefont
  {Bart{\'o}k}}, \bibinfo {author} {\bibfnamefont {M.~C.}\ \bibnamefont
  {Payne}}, \bibinfo {author} {\bibfnamefont {R.}~\bibnamefont {Kondor}},\ and\
  \bibinfo {author} {\bibfnamefont {G.}~\bibnamefont {Cs{\'a}nyi}},\ }\bibfield
   {title} {\enquote {\bibinfo {title} {Gaussian {{Approximation Potentials}}:
  {{The Accuracy}} of {{Quantum Mechanics}}, without the {{Electrons}}},}\
  }\href {https://doi.org/10.1103/PhysRevLett.104.136403} {\bibfield  {journal}
  {\bibinfo  {journal} {\authorjournal{Phys. Rev. Lett.}{Physical Review
  Letters}}\ }\textbf {\bibinfo {volume} {104}},\ \bibinfo {pages} {136403}
  (\bibinfo {year} {2010})}\BibitemShut {NoStop}%
\bibitem [{\citenamefont {Bart{\'o}k}, \citenamefont {Kondor},\ and\
  \citenamefont
  {Cs{\'a}nyi}(2013)}]{bartokRepresentingChemicalEnvironments2013}%
  \BibitemOpen
  \bibfield  {author} {\bibinfo {author} {\bibfnamefont {A.~P.}\ \bibnamefont
  {Bart{\'o}k}}, \bibinfo {author} {\bibfnamefont {R.}~\bibnamefont {Kondor}},\
  and\ \bibinfo {author} {\bibfnamefont {G.}~\bibnamefont {Cs{\'a}nyi}},\
  }\bibfield  {title} {\enquote {\bibinfo {title} {On representing chemical
  environments},}\ }\href {https://doi.org/10.1103/PhysRevB.87.184115}
  {\bibfield  {journal} {\bibinfo  {journal} {\authorjournal{Phys. Rev.
  B}{Physical Review B}}\ }\textbf {\bibinfo {volume} {87}},\ \bibinfo {pages}
  {184115} (\bibinfo {year} {2013})}\BibitemShut {NoStop}%
\bibitem [{\citenamefont {Bart{\'o}k}\ \emph {et~al.}(2018)\citenamefont
  {Bart{\'o}k}, \citenamefont {Kermode}, \citenamefont {Bernstein},\ and\
  \citenamefont {Cs{\'a}nyi}}]{bartokMachineLearningGeneralPurpose2018}%
  \BibitemOpen
  \bibfield  {author} {\bibinfo {author} {\bibfnamefont {A.~P.}\ \bibnamefont
  {Bart{\'o}k}}, \bibinfo {author} {\bibfnamefont {J.}~\bibnamefont {Kermode}},
  \bibinfo {author} {\bibfnamefont {N.}~\bibnamefont {Bernstein}},\ and\
  \bibinfo {author} {\bibfnamefont {G.}~\bibnamefont {Cs{\'a}nyi}},\ }\bibfield
   {title} {\enquote {\bibinfo {title} {Machine {{Learning}} a
  {{General-Purpose Interatomic Potential}} for {{Silicon}}},}\ }\href
  {https://doi.org/10.1103/PhysRevX.8.041048} {\bibfield  {journal} {\bibinfo
  {journal} {\authorjournal{Phys. Rev. X}{Physical Review X}}\ }\textbf
  {\bibinfo {volume} {8}},\ \bibinfo {pages} {041048} (\bibinfo {year}
  {2018})}\BibitemShut {NoStop}%
\bibitem [{\citenamefont {Thompson}\ \emph {et~al.}(2015)\citenamefont
  {Thompson}, \citenamefont {Swiler}, \citenamefont {Trott}, \citenamefont
  {Foiles},\ and\ \citenamefont
  {Tucker}}]{thompsonSpectralNeighborAnalysis2015a}%
  \BibitemOpen
  \bibfield  {author} {\bibinfo {author} {\bibfnamefont {A.~P.}\ \bibnamefont
  {Thompson}}, \bibinfo {author} {\bibfnamefont {L.~P.}\ \bibnamefont
  {Swiler}}, \bibinfo {author} {\bibfnamefont {C.~R.}\ \bibnamefont {Trott}},
  \bibinfo {author} {\bibfnamefont {S.~M.}\ \bibnamefont {Foiles}},\ and\
  \bibinfo {author} {\bibfnamefont {G.~J.}\ \bibnamefont {Tucker}},\ }\bibfield
   {title} {\enquote {\bibinfo {title} {Spectral neighbor analysis method for
  automated generation of quantum-accurate interatomic potentials},}\ }\href
  {https://doi.org/10.1016/j.jcp.2014.12.018} {\bibfield  {journal} {\bibinfo
  {journal} {\authorjournal{J. Comput. Phys.}{Journal of Computational
  Physics}}\ }\textbf {\bibinfo {volume} {285}},\ \bibinfo {pages} {316--330}
  (\bibinfo {year} {2015})}\BibitemShut {NoStop}%
\bibitem [{\citenamefont {Shapeev}(2016)}]{shapeevMomentTensorPotentials2016}%
  \BibitemOpen
  \bibfield  {author} {\bibinfo {author} {\bibfnamefont {A.~V.}\ \bibnamefont
  {Shapeev}},\ }\bibfield  {title} {\enquote {\bibinfo {title} {Moment {{Tensor
  Potentials}}: {{A Class}} of {{Systematically Improvable Interatomic
  Potentials}}},}\ }\href {https://doi.org/10.1137/15M1054183} {\bibfield
  {journal} {\bibinfo  {journal} {\authorjournal{Multiscale Model.
  Simul.}{Multiscale Modeling \& Simulation}}\ }\textbf {\bibinfo {volume}
  {14}},\ \bibinfo {pages} {1153--1173} (\bibinfo {year} {2016})}\BibitemShut
  {NoStop}%
\bibitem [{\citenamefont {Zhang}\ \emph {et~al.}(2018)\citenamefont {Zhang},
  \citenamefont {Han}, \citenamefont {Wang}, \citenamefont {Car},\ and\
  \citenamefont {E}}]{zhangDeepPotentialMolecular2018}%
  \BibitemOpen
  \bibfield  {author} {\bibinfo {author} {\bibfnamefont {L.}~\bibnamefont
  {Zhang}}, \bibinfo {author} {\bibfnamefont {J.}~\bibnamefont {Han}}, \bibinfo
  {author} {\bibfnamefont {H.}~\bibnamefont {Wang}}, \bibinfo {author}
  {\bibfnamefont {R.}~\bibnamefont {Car}},\ and\ \bibinfo {author}
  {\bibfnamefont {W.}~\bibnamefont {E}},\ }\bibfield  {title} {\enquote
  {\bibinfo {title} {Deep {{Potential Molecular Dynamics}}: {{A Scalable
  Model}} with the {{Accuracy}} of {{Quantum Mechanics}}},}\ }\href
  {https://doi.org/10.1103/PhysRevLett.120.143001} {\bibfield  {journal}
  {\bibinfo  {journal} {\authorjournal{Phys. Rev. Lett.}{Physical Review
  Letters}}\ }\textbf {\bibinfo {volume} {120}},\ \bibinfo {pages} {143001}
  (\bibinfo {year} {2018})}\BibitemShut {NoStop}%
\bibitem [{\citenamefont {Batzner}\ \emph {et~al.}(2022)\citenamefont
  {Batzner}, \citenamefont {Musaelian}, \citenamefont {Sun}, \citenamefont
  {Geiger}, \citenamefont {Mailoa}, \citenamefont {Kornbluth}, \citenamefont
  {Molinari}, \citenamefont {Smidt},\ and\ \citenamefont
  {Kozinsky}}]{batznerE3equivariantGraphNeural2022}%
  \BibitemOpen
  \bibfield  {author} {\bibinfo {author} {\bibfnamefont {S.}~\bibnamefont
  {Batzner}}, \bibinfo {author} {\bibfnamefont {A.}~\bibnamefont {Musaelian}},
  \bibinfo {author} {\bibfnamefont {L.}~\bibnamefont {Sun}}, \bibinfo {author}
  {\bibfnamefont {M.}~\bibnamefont {Geiger}}, \bibinfo {author} {\bibfnamefont
  {J.~P.}\ \bibnamefont {Mailoa}}, \bibinfo {author} {\bibfnamefont
  {M.}~\bibnamefont {Kornbluth}}, \bibinfo {author} {\bibfnamefont
  {N.}~\bibnamefont {Molinari}}, \bibinfo {author} {\bibfnamefont {T.~E.}\
  \bibnamefont {Smidt}},\ and\ \bibinfo {author} {\bibfnamefont
  {B.}~\bibnamefont {Kozinsky}},\ }\bibfield  {title} {\enquote {\bibinfo
  {title} {E(3)-equivariant graph neural networks for data-efficient and
  accurate interatomic potentials},}\ }\href
  {https://doi.org/10.1038/s41467-022-29939-5} {\bibfield  {journal} {\bibinfo
  {journal} {\authorjournal{Nat. Commun.}{Nature Communications}}\ }\textbf
  {\bibinfo {volume} {13}},\ \bibinfo {pages} {2453} (\bibinfo {year}
  {2022})}\BibitemShut {NoStop}%
\bibitem [{\citenamefont {Batatia}\ \emph {et~al.}(2022)\citenamefont
  {Batatia}, \citenamefont {Kovacs}, \citenamefont {Simm}, \citenamefont
  {Ortner},\ and\ \citenamefont {Csanyi}}]{batatiaMACEHigherOrder2022}%
  \BibitemOpen
  \bibfield  {author} {\bibinfo {author} {\bibfnamefont {I.}~\bibnamefont
  {Batatia}}, \bibinfo {author} {\bibfnamefont {D.~P.}\ \bibnamefont {Kovacs}},
  \bibinfo {author} {\bibfnamefont {G.}~\bibnamefont {Simm}}, \bibinfo {author}
  {\bibfnamefont {C.}~\bibnamefont {Ortner}},\ and\ \bibinfo {author}
  {\bibfnamefont {G.}~\bibnamefont {Csanyi}},\ }\bibfield  {title} {\enquote
  {\bibinfo {title} {{{MACE}}: {{Higher Order Equivariant Message Passing
  Neural Networks}} for {{Fast}} and {{Accurate Force Fields}}},}\ }in\ \href
  {https://doi.org/10.52202/068431-0830} {\emph {\bibinfo {booktitle} {Advances
  in {{Neural Information Processing Systems}} 35}}}\ (\bibinfo  {publisher}
  {Neural Information Processing Systems Foundation, Inc. (NeurIPS)},\ \bibinfo
  {address} {New Orleans, Louisiana, USA},\ \bibinfo {year} {2022})\ pp.\
  \bibinfo {pages} {11423--11436}\BibitemShut {NoStop}%
\bibitem [{\citenamefont {Leimeroth}\ \emph {et~al.}(2025)\citenamefont
  {Leimeroth}, \citenamefont {Erhard}, \citenamefont {Albe},\ and\
  \citenamefont {Rohrer}}]{leimerothMachinelearningInteratomicPotentials2025}%
  \BibitemOpen
  \bibfield  {author} {\bibinfo {author} {\bibfnamefont {N.}~\bibnamefont
  {Leimeroth}}, \bibinfo {author} {\bibfnamefont {L.~C.}\ \bibnamefont
  {Erhard}}, \bibinfo {author} {\bibfnamefont {K.}~\bibnamefont {Albe}},\ and\
  \bibinfo {author} {\bibfnamefont {J.}~\bibnamefont {Rohrer}},\ }\bibfield
  {title} {\enquote {\bibinfo {title} {Machine-learning interatomic potentials
  from a users perspective: {{A}} comparison of accuracy, speed and data
  efficiency},}\ }\href {https://doi.org/10.1088/1361-651X/adf56d} {\bibfield
  {journal} {\bibinfo  {journal} {\authorjournal{Modell. Simul. Mater. Sci.
  Eng.}{Modelling and Simulation in Materials Science and Engineering}}\
  }\textbf {\bibinfo {volume} {33}},\ \bibinfo {pages} {065012} (\bibinfo
  {year} {2025})},\ \Eprint {https://arxiv.org/abs/2505.02503}
  {arXiv:2505.02503 [cond-mat.mtrl-sci]} \BibitemShut {NoStop}%
\bibitem [{\citenamefont {Xue}\ \emph {et~al.}(2021)\citenamefont {Xue},
  \citenamefont {Guo}, \citenamefont {Wen}, \citenamefont {Feng}, \citenamefont
  {Huang}, \citenamefont {Guo}, \citenamefont {Li}, \citenamefont {Cui},
  \citenamefont {Zhang},\ and\ \citenamefont
  {Wang}}]{xueReaxFFMPNNMachineLearning2021}%
  \BibitemOpen
  \bibfield  {author} {\bibinfo {author} {\bibfnamefont {L.-Y.}\ \bibnamefont
  {Xue}}, \bibinfo {author} {\bibfnamefont {F.}~\bibnamefont {Guo}}, \bibinfo
  {author} {\bibfnamefont {Y.-S.}\ \bibnamefont {Wen}}, \bibinfo {author}
  {\bibfnamefont {S.-Q.}\ \bibnamefont {Feng}}, \bibinfo {author}
  {\bibfnamefont {X.-N.}\ \bibnamefont {Huang}}, \bibinfo {author}
  {\bibfnamefont {L.}~\bibnamefont {Guo}}, \bibinfo {author} {\bibfnamefont
  {H.-S.}\ \bibnamefont {Li}}, \bibinfo {author} {\bibfnamefont {S.-X.}\
  \bibnamefont {Cui}}, \bibinfo {author} {\bibfnamefont {G.-Q.}\ \bibnamefont
  {Zhang}},\ and\ \bibinfo {author} {\bibfnamefont {Q.-L.}\ \bibnamefont
  {Wang}},\ }\bibfield  {title} {\enquote {\bibinfo {title} {{{ReaxFF-MPNN}}
  machine learning potential: A combination of reactive force field and message
  passing neural networks},}\ }\href {https://doi.org/10.1039/D1CP01656C}
  {\bibfield  {journal} {\bibinfo  {journal} {\authorjournal{Phys. Chem. Chem.
  Phys.}{Physical Chemistry Chemical Physics}}\ }\textbf {\bibinfo {volume}
  {23}},\ \bibinfo {pages} {19457--19464} (\bibinfo {year} {2021})}\BibitemShut
  {NoStop}%
\bibitem [{\citenamefont {Ye}\ \emph {et~al.}(2025)\citenamefont {Ye},
  \citenamefont {Liu}, \citenamefont {Chai}, \citenamefont {Wen}, \citenamefont
  {Cui}, \citenamefont {Zhang}, \citenamefont {Li}, \citenamefont {Guo},\ and\
  \citenamefont {Wang}}]{yeReaxFFnnReactiveMachinelearning2025}%
  \BibitemOpen
  \bibfield  {author} {\bibinfo {author} {\bibfnamefont {Z.-H.}\ \bibnamefont
  {Ye}}, \bibinfo {author} {\bibfnamefont {J.-H.}\ \bibnamefont {Liu}},
  \bibinfo {author} {\bibfnamefont {C.-G.}\ \bibnamefont {Chai}}, \bibinfo
  {author} {\bibfnamefont {Y.-S.}\ \bibnamefont {Wen}}, \bibinfo {author}
  {\bibfnamefont {S.-X.}\ \bibnamefont {Cui}}, \bibinfo {author} {\bibfnamefont
  {G.-Q.}\ \bibnamefont {Zhang}}, \bibinfo {author} {\bibfnamefont {K.-J.}\
  \bibnamefont {Li}}, \bibinfo {author} {\bibfnamefont {F.}~\bibnamefont
  {Guo}},\ and\ \bibinfo {author} {\bibfnamefont {X.-C.}\ \bibnamefont
  {Wang}},\ }\bibfield  {title} {\enquote {\bibinfo {title} {{{ReaxFF}}-nn: A
  reactive machine-learning potential in {{GULP}}/{{LAMMPS}} and its
  applications in the thermal conductivity calculations of carbon
  nanostructures},}\ }\href {https://doi.org/10.1039/D4CP00535J} {\bibfield
  {journal} {\bibinfo  {journal} {\authorjournal{Phys. Chem. Chem.
  Phys.}{Physical Chemistry Chemical Physics}}\ }\textbf {\bibinfo {volume}
  {27}},\ \bibinfo {pages} {10571--10579} (\bibinfo {year} {2025})}\BibitemShut
  {NoStop}%
\bibitem [{\citenamefont
  {Abell}(1985)}]{abellEmpiricalChemicalPseudopotential1985}%
  \BibitemOpen
  \bibfield  {author} {\bibinfo {author} {\bibfnamefont {G.~C.}\ \bibnamefont
  {Abell}},\ }\bibfield  {title} {\enquote {\bibinfo {title} {Empirical
  chemical pseudopotential theory of molecular and metallic bonding},}\ }\href
  {https://doi.org/10.1103/PhysRevB.31.6184} {\bibfield  {journal} {\bibinfo
  {journal} {\authorjournal{Phys. Rev. B}{Physical Review B}}\ }\textbf
  {\bibinfo {volume} {31}},\ \bibinfo {pages} {6184--6196} (\bibinfo {year}
  {1985})}\BibitemShut {NoStop}%
\bibitem [{\citenamefont
  {Tersoff}(1988{\natexlab{a}})}]{tersoffNewEmpiricalApproach1988}%
  \BibitemOpen
  \bibfield  {author} {\bibinfo {author} {\bibfnamefont {J.}~\bibnamefont
  {Tersoff}},\ }\bibfield  {title} {\enquote {\bibinfo {title} {New empirical
  approach for the structure and energy of covalent systems},}\ }\href
  {https://doi.org/10.1103/PhysRevB.37.6991} {\bibfield  {journal} {\bibinfo
  {journal} {\authorjournal{Phys. Rev. B}{Physical Review B}}\ }\textbf
  {\bibinfo {volume} {37}},\ \bibinfo {pages} {6991--7000} (\bibinfo {year}
  {1988}{\natexlab{a}})}\BibitemShut {NoStop}%
\bibitem [{\citenamefont {Hobday}, \citenamefont {Smith},\ and\ \citenamefont
  {Belbruno}(1999)}]{hobdayApplicationsNeuralNetworks1999}%
  \BibitemOpen
  \bibfield  {author} {\bibinfo {author} {\bibfnamefont {S.}~\bibnamefont
  {Hobday}}, \bibinfo {author} {\bibfnamefont {R.}~\bibnamefont {Smith}},\ and\
  \bibinfo {author} {\bibfnamefont {J.}~\bibnamefont {Belbruno}},\ }\bibfield
  {title} {\enquote {\bibinfo {title} {Applications of neural networks to
  fitting interatomic potential functions},}\ }\href
  {https://doi.org/10.1088/0965-0393/7/3/308} {\bibfield  {journal} {\bibinfo
  {journal} {\authorjournal{Modell. Simul. Mater. Sci. Eng.}{Modelling and
  Simulation in Materials Science and Engineering}}\ }\textbf {\bibinfo
  {volume} {7}},\ \bibinfo {pages} {397--412} (\bibinfo {year}
  {1999})}\BibitemShut {NoStop}%
\bibitem [{\citenamefont {Malshe}\ \emph {et~al.}(2008)\citenamefont {Malshe},
  \citenamefont {Narulkar}, \citenamefont {Raff}, \citenamefont {Hagan},
  \citenamefont {Bukkapatnam},\ and\ \citenamefont
  {Komanduri}}]{malsheParametrizationAnalyticInteratomic2008}%
  \BibitemOpen
  \bibfield  {author} {\bibinfo {author} {\bibfnamefont {M.}~\bibnamefont
  {Malshe}}, \bibinfo {author} {\bibfnamefont {R.}~\bibnamefont {Narulkar}},
  \bibinfo {author} {\bibfnamefont {L.~M.}\ \bibnamefont {Raff}}, \bibinfo
  {author} {\bibfnamefont {M.}~\bibnamefont {Hagan}}, \bibinfo {author}
  {\bibfnamefont {S.}~\bibnamefont {Bukkapatnam}},\ and\ \bibinfo {author}
  {\bibfnamefont {R.}~\bibnamefont {Komanduri}},\ }\bibfield  {title} {\enquote
  {\bibinfo {title} {Parametrization of analytic interatomic potential
  functions using neural networks},}\ }\href
  {https://doi.org/10.1063/1.2957490} {\bibfield  {journal} {\bibinfo
  {journal} {\authorjournal{J. Chem. Phys.}{The Journal of Chemical Physics}}\
  }\textbf {\bibinfo {volume} {129}},\ \bibinfo {pages} {044111} (\bibinfo
  {year} {2008})}\BibitemShut {NoStop}%
\bibitem [{\citenamefont {Pun}\ \emph {et~al.}(2019)\citenamefont {Pun},
  \citenamefont {Batra}, \citenamefont {Ramprasad},\ and\ \citenamefont
  {Mishin}}]{punPhysicallyInformedArtificial2019}%
  \BibitemOpen
  \bibfield  {author} {\bibinfo {author} {\bibfnamefont {G.~P.~P.}\
  \bibnamefont {Pun}}, \bibinfo {author} {\bibfnamefont {R.}~\bibnamefont
  {Batra}}, \bibinfo {author} {\bibfnamefont {R.}~\bibnamefont {Ramprasad}},\
  and\ \bibinfo {author} {\bibfnamefont {Y.}~\bibnamefont {Mishin}},\
  }\bibfield  {title} {\enquote {\bibinfo {title} {Physically informed
  artificial neural networks for atomistic modeling of materials},}\ }\href
  {https://doi.org/10.1038/s41467-019-10343-5} {\bibfield  {journal} {\bibinfo
  {journal} {\authorjournal{Nat. Commun.}{Nature Communications}}\ }\textbf
  {\bibinfo {volume} {10}},\ \bibinfo {pages} {2339} (\bibinfo {year}
  {2019})}\BibitemShut {NoStop}%
\bibitem [{\citenamefont {Kohata}\ \emph {et~al.}(2025)\citenamefont {Kohata},
  \citenamefont {Hisama}, \citenamefont {Otsuka},\ and\ \citenamefont
  {Maruyama}}]{kohataMachineLearnedBondOrderPotential2025}%
  \BibitemOpen
  \bibfield  {author} {\bibinfo {author} {\bibfnamefont {I.}~\bibnamefont
  {Kohata}}, \bibinfo {author} {\bibfnamefont {K.}~\bibnamefont {Hisama}},
  \bibinfo {author} {\bibfnamefont {K.}~\bibnamefont {Otsuka}},\ and\ \bibinfo
  {author} {\bibfnamefont {S.}~\bibnamefont {Maruyama}},\ }\bibfield  {title}
  {\enquote {\bibinfo {title} {Machine-{{Learned Bond-Order Potential}} for
  {{Exploring}} the {{Configuration Space}} of {{Carbon}}},}\ }\href
  {https://doi.org/10.1103/73gp-bm46} {\bibfield  {journal} {\bibinfo
  {journal} {\authorjournal{Phys. Rev. Mater.}{Physical Review Materials}}\
  }\textbf {\bibinfo {volume} {9}},\ \bibinfo {pages} {074009} (\bibinfo {year}
  {2025})},\ \Eprint {https://arxiv.org/abs/2501.11297} {arXiv:2501.11297
  [cond-mat.mtrl-sci]} \BibitemShut {NoStop}%
\bibitem [{\citenamefont
  {Tersoff}(1988{\natexlab{b}})}]{tersoffEmpiricalInteratomicPotential1988}%
  \BibitemOpen
  \bibfield  {author} {\bibinfo {author} {\bibfnamefont {J.}~\bibnamefont
  {Tersoff}},\ }\bibfield  {title} {\enquote {\bibinfo {title} {Empirical
  interatomic potential for silicon with improved elastic properties},}\ }\href
  {https://doi.org/10.1103/PhysRevB.38.9902} {\bibfield  {journal} {\bibinfo
  {journal} {\authorjournal{Phys. Rev. B}{Physical Review B}}\ }\textbf
  {\bibinfo {volume} {38}},\ \bibinfo {pages} {9902--9905} (\bibinfo {year}
  {1988}{\natexlab{b}})}\BibitemShut {NoStop}%
\bibitem [{\citenamefont
  {Tersoff}(1989)}]{tersoffModelingSolidstateChemistry1989}%
  \BibitemOpen
  \bibfield  {author} {\bibinfo {author} {\bibfnamefont {J.}~\bibnamefont
  {Tersoff}},\ }\bibfield  {title} {\enquote {\bibinfo {title} {Modeling
  solid-state chemistry: {{Interatomic}} potentials for multicomponent
  systems},}\ }\href {https://doi.org/10.1103/PhysRevB.39.5566} {\bibfield
  {journal} {\bibinfo  {journal} {\authorjournal{Phys. Rev. B}{Physical Review
  B}}\ }\textbf {\bibinfo {volume} {39}},\ \bibinfo {pages} {5566--5568}
  (\bibinfo {year} {1989})}\BibitemShut {NoStop}%
\bibitem [{\citenamefont {Sch{\"u}tt}\ \emph {et~al.}(2017)\citenamefont
  {Sch{\"u}tt}, \citenamefont {Arbabzadah}, \citenamefont {Chmiela},
  \citenamefont {M{\"u}ller},\ and\ \citenamefont
  {Tkatchenko}}]{schuttQuantumchemicalInsightsDeep2017}%
  \BibitemOpen
  \bibfield  {author} {\bibinfo {author} {\bibfnamefont {K.~T.}\ \bibnamefont
  {Sch{\"u}tt}}, \bibinfo {author} {\bibfnamefont {F.}~\bibnamefont
  {Arbabzadah}}, \bibinfo {author} {\bibfnamefont {S.}~\bibnamefont {Chmiela}},
  \bibinfo {author} {\bibfnamefont {K.~R.}\ \bibnamefont {M{\"u}ller}},\ and\
  \bibinfo {author} {\bibfnamefont {A.}~\bibnamefont {Tkatchenko}},\ }\bibfield
   {title} {\enquote {\bibinfo {title} {Quantum-chemical insights from deep
  tensor neural networks},}\ }\href {https://doi.org/10.1038/ncomms13890}
  {\bibfield  {journal} {\bibinfo  {journal} {\authorjournal{Nat.
  Commun.}{Nature Communications}}\ }\textbf {\bibinfo {volume} {8}},\ \bibinfo
  {pages} {13890} (\bibinfo {year} {2017})}\BibitemShut {NoStop}%
\bibitem [{\citenamefont {Sch{\"u}tt}\ \emph {et~al.}(2018)\citenamefont
  {Sch{\"u}tt}, \citenamefont {Sauceda}, \citenamefont {Kindermans},
  \citenamefont {Tkatchenko},\ and\ \citenamefont
  {M{\"u}ller}}]{schuttSchNetDeepLearning2018}%
  \BibitemOpen
  \bibfield  {author} {\bibinfo {author} {\bibfnamefont {K.~T.}\ \bibnamefont
  {Sch{\"u}tt}}, \bibinfo {author} {\bibfnamefont {H.~E.}\ \bibnamefont
  {Sauceda}}, \bibinfo {author} {\bibfnamefont {P.-J.}\ \bibnamefont
  {Kindermans}}, \bibinfo {author} {\bibfnamefont {A.}~\bibnamefont
  {Tkatchenko}},\ and\ \bibinfo {author} {\bibfnamefont {K.-R.}\ \bibnamefont
  {M{\"u}ller}},\ }\bibfield  {title} {\enquote {\bibinfo {title} {{{SchNet}}
  -- {{A}} deep learning architecture for molecules and materials},}\ }\href
  {https://doi.org/10.1063/1.5019779} {\bibfield  {journal} {\bibinfo
  {journal} {\authorjournal{J. Chem. Phys.}{The Journal of Chemical Physics}}\
  }\textbf {\bibinfo {volume} {148}},\ \bibinfo {pages} {241722} (\bibinfo
  {year} {2018})}\BibitemShut {NoStop}%
\bibitem [{\citenamefont {Dugas}\ \emph {et~al.}(2000)\citenamefont {Dugas},
  \citenamefont {Bengio}, \citenamefont {B{\'e}lisle}, \citenamefont {Nadeau},\
  and\ \citenamefont {Garcia}}]{dugasIncorporatingSecondorderFunctional2000}%
  \BibitemOpen
  \bibfield  {author} {\bibinfo {author} {\bibfnamefont {C.}~\bibnamefont
  {Dugas}}, \bibinfo {author} {\bibfnamefont {Y.}~\bibnamefont {Bengio}},
  \bibinfo {author} {\bibfnamefont {F.}~\bibnamefont {B{\'e}lisle}}, \bibinfo
  {author} {\bibfnamefont {C.}~\bibnamefont {Nadeau}},\ and\ \bibinfo {author}
  {\bibfnamefont {R.}~\bibnamefont {Garcia}},\ }\bibfield  {title} {\enquote
  {\bibinfo {title} {Incorporating second-order functional knowledge for better
  option pricing},}\ }in\ \href@noop {} {\emph {\bibinfo {booktitle} {Advances
  in {{Neural Information Processing Systems}}}}},\ Vol.~\bibinfo {volume}
  {13}\ (\bibinfo  {publisher} {MIT Press},\ \bibinfo {year}
  {2000})\BibitemShut {NoStop}%
\bibitem [{\citenamefont {Daniels}\ and\ \citenamefont
  {Velikova}(2010)}]{danielsMonotonePartiallyMonotone2010}%
  \BibitemOpen
  \bibfield  {author} {\bibinfo {author} {\bibfnamefont {H.}~\bibnamefont
  {Daniels}}\ and\ \bibinfo {author} {\bibfnamefont {M.}~\bibnamefont
  {Velikova}},\ }\bibfield  {title} {\enquote {\bibinfo {title} {Monotone and
  partially monotone neural networks},}\ }\href
  {https://doi.org/10.1109/TNN.2010.2044803} {\bibfield  {journal} {\bibinfo
  {journal} {\authorjournal{IEEE Trans. Neural Netw.}{IEEE Transactions on
  Neural Networks}}\ }\textbf {\bibinfo {volume} {21}},\ \bibinfo {pages}
  {906--917} (\bibinfo {year} {2010})}\BibitemShut {NoStop}%
\bibitem [{\citenamefont {Runje}\ and\ \citenamefont
  {Shankaranarayana}(2023)}]{runjeConstrainedMonotonicNeural2023}%
  \BibitemOpen
  \bibfield  {author} {\bibinfo {author} {\bibfnamefont {D.}~\bibnamefont
  {Runje}}\ and\ \bibinfo {author} {\bibfnamefont {S.~M.}\ \bibnamefont
  {Shankaranarayana}},\ }\bibfield  {title} {\enquote {\bibinfo {title}
  {Constrained monotonic neural networks},}\ }in\ \href@noop {} {\emph
  {\bibinfo {booktitle} {Proceedings of the 40th {{International Conference}}
  on {{Machine Learning}}}}}\ (\bibinfo  {publisher} {PMLR},\ \bibinfo {year}
  {2023})\ pp.\ \bibinfo {pages} {29338--29353}\BibitemShut {NoStop}%
\bibitem [{\citenamefont {Elfwing}, \citenamefont {Uchibe},\ and\ \citenamefont
  {Doya}(2018)}]{elfwingSigmoidweightedLinearUnits2018}%
  \BibitemOpen
  \bibfield  {author} {\bibinfo {author} {\bibfnamefont {S.}~\bibnamefont
  {Elfwing}}, \bibinfo {author} {\bibfnamefont {E.}~\bibnamefont {Uchibe}},\
  and\ \bibinfo {author} {\bibfnamefont {K.}~\bibnamefont {Doya}},\ }\bibfield
  {title} {\enquote {\bibinfo {title} {Sigmoid-weighted linear units for neural
  network function approximation in reinforcement learning},}\ }\href
  {https://doi.org/10.1016/j.neunet.2017.12.012} {\bibfield  {journal}
  {\bibinfo  {journal} {\authorjournal{Neural Netw.}{Neural Networks}}\
  }\bibinfo {series} {Special Issue on Deep Reinforcement Learning},\ \textbf
  {\bibinfo {volume} {107}},\ \bibinfo {pages} {3--11} (\bibinfo {year}
  {2018})}\BibitemShut {NoStop}%
\bibitem [{\citenamefont {Kresse}\ and\ \citenamefont
  {Furthm{\"u}ller}(1996)}]{kresseEfficientIterativeSchemes1996}%
  \BibitemOpen
  \bibfield  {author} {\bibinfo {author} {\bibfnamefont {G.}~\bibnamefont
  {Kresse}}\ and\ \bibinfo {author} {\bibfnamefont {J.}~\bibnamefont
  {Furthm{\"u}ller}},\ }\bibfield  {title} {\enquote {\bibinfo {title}
  {Efficient iterative schemes for {\emph{ab initio}} total-energy calculations
  using a plane-wave basis set},}\ }\href
  {https://doi.org/10.1103/PhysRevB.54.11169} {\bibfield  {journal} {\bibinfo
  {journal} {\authorjournal{Phys. Rev. B}{Physical Review B}}\ }\textbf
  {\bibinfo {volume} {54}},\ \bibinfo {pages} {11169--11186} (\bibinfo {year}
  {1996})}\BibitemShut {NoStop}%
\bibitem [{\citenamefont {Kresse}\ and\ \citenamefont
  {Joubert}(1999)}]{kresseUltrasoftPseudopotentialsProjector1999}%
  \BibitemOpen
  \bibfield  {author} {\bibinfo {author} {\bibfnamefont {G.}~\bibnamefont
  {Kresse}}\ and\ \bibinfo {author} {\bibfnamefont {D.}~\bibnamefont
  {Joubert}},\ }\bibfield  {title} {\enquote {\bibinfo {title} {From ultrasoft
  pseudopotentials to the projector augmented-wave method},}\ }\href
  {https://doi.org/10.1103/PhysRevB.59.1758} {\bibfield  {journal} {\bibinfo
  {journal} {\authorjournal{Phys. Rev. B}{Physical Review B}}\ }\textbf
  {\bibinfo {volume} {59}},\ \bibinfo {pages} {1758--1775} (\bibinfo {year}
  {1999})}\BibitemShut {NoStop}%
\bibitem [{\citenamefont {Perdew}, \citenamefont {Burke},\ and\ \citenamefont
  {Ernzerhof}(1996)}]{perdewGeneralizedGradientApproximation1996}%
  \BibitemOpen
  \bibfield  {author} {\bibinfo {author} {\bibfnamefont {J.~P.}\ \bibnamefont
  {Perdew}}, \bibinfo {author} {\bibfnamefont {K.}~\bibnamefont {Burke}},\ and\
  \bibinfo {author} {\bibfnamefont {M.}~\bibnamefont {Ernzerhof}},\ }\bibfield
  {title} {\enquote {\bibinfo {title} {Generalized {{Gradient Approximation
  Made Simple}}},}\ }\href {https://doi.org/10.1103/PhysRevLett.77.3865}
  {\bibfield  {journal} {\bibinfo  {journal} {\authorjournal{Phys. Rev.
  Lett.}{Physical Review Letters}}\ }\textbf {\bibinfo {volume} {77}},\
  \bibinfo {pages} {3865--3868} (\bibinfo {year} {1996})}\BibitemShut {NoStop}%
\bibitem [{\citenamefont {Hjorth~Larsen}\ \emph {et~al.}(2017)\citenamefont
  {Hjorth~Larsen}, \citenamefont {J{\o}rgen~Mortensen}, \citenamefont
  {Blomqvist}, \citenamefont {Castelli}, \citenamefont {Christensen},
  \citenamefont {Du{\l}ak}, \citenamefont {Friis}, \citenamefont {Groves},
  \citenamefont {Hammer}, \citenamefont {Hargus}, \citenamefont {Hermes},
  \citenamefont {Jennings}, \citenamefont {Bjerre~Jensen}, \citenamefont
  {Kermode}, \citenamefont {Kitchin}, \citenamefont {Leonhard~Kolsbjerg},
  \citenamefont {Kubal}, \citenamefont {Kaasbjerg}, \citenamefont {Lysgaard},
  \citenamefont {Bergmann~Maronsson}, \citenamefont {Maxson}, \citenamefont
  {Olsen}, \citenamefont {Pastewka}, \citenamefont {Peterson}, \citenamefont
  {Rostgaard}, \citenamefont {Schi{\o}tz}, \citenamefont {Sch{\"u}tt},
  \citenamefont {Strange}, \citenamefont {Thygesen}, \citenamefont {Vegge},
  \citenamefont {Vilhelmsen}, \citenamefont {Walter}, \citenamefont {Zeng},\
  and\ \citenamefont {Jacobsen}}]{hjorthlarsenAtomicSimulationEnvironment2017}%
  \BibitemOpen
  \bibfield  {author} {\bibinfo {author} {\bibfnamefont {A.}~\bibnamefont
  {Hjorth~Larsen}}, \bibinfo {author} {\bibfnamefont {J.}~\bibnamefont
  {J{\o}rgen~Mortensen}}, \bibinfo {author} {\bibfnamefont {J.}~\bibnamefont
  {Blomqvist}}, \bibinfo {author} {\bibfnamefont {I.~E.}\ \bibnamefont
  {Castelli}}, \bibinfo {author} {\bibfnamefont {R.}~\bibnamefont
  {Christensen}}, \bibinfo {author} {\bibfnamefont {M.}~\bibnamefont
  {Du{\l}ak}}, \bibinfo {author} {\bibfnamefont {J.}~\bibnamefont {Friis}},
  \bibinfo {author} {\bibfnamefont {M.~N.}\ \bibnamefont {Groves}}, \bibinfo
  {author} {\bibfnamefont {B.}~\bibnamefont {Hammer}}, \bibinfo {author}
  {\bibfnamefont {C.}~\bibnamefont {Hargus}}, \bibinfo {author} {\bibfnamefont
  {E.~D.}\ \bibnamefont {Hermes}}, \bibinfo {author} {\bibfnamefont {P.~C.}\
  \bibnamefont {Jennings}}, \bibinfo {author} {\bibfnamefont {P.}~\bibnamefont
  {Bjerre~Jensen}}, \bibinfo {author} {\bibfnamefont {J.}~\bibnamefont
  {Kermode}}, \bibinfo {author} {\bibfnamefont {J.~R.}\ \bibnamefont
  {Kitchin}}, \bibinfo {author} {\bibfnamefont {E.}~\bibnamefont
  {Leonhard~Kolsbjerg}}, \bibinfo {author} {\bibfnamefont {J.}~\bibnamefont
  {Kubal}}, \bibinfo {author} {\bibfnamefont {K.}~\bibnamefont {Kaasbjerg}},
  \bibinfo {author} {\bibfnamefont {S.}~\bibnamefont {Lysgaard}}, \bibinfo
  {author} {\bibfnamefont {J.}~\bibnamefont {Bergmann~Maronsson}}, \bibinfo
  {author} {\bibfnamefont {T.}~\bibnamefont {Maxson}}, \bibinfo {author}
  {\bibfnamefont {T.}~\bibnamefont {Olsen}}, \bibinfo {author} {\bibfnamefont
  {L.}~\bibnamefont {Pastewka}}, \bibinfo {author} {\bibfnamefont
  {A.}~\bibnamefont {Peterson}}, \bibinfo {author} {\bibfnamefont
  {C.}~\bibnamefont {Rostgaard}}, \bibinfo {author} {\bibfnamefont
  {J.}~\bibnamefont {Schi{\o}tz}}, \bibinfo {author} {\bibfnamefont
  {O.}~\bibnamefont {Sch{\"u}tt}}, \bibinfo {author} {\bibfnamefont
  {M.}~\bibnamefont {Strange}}, \bibinfo {author} {\bibfnamefont {K.~S.}\
  \bibnamefont {Thygesen}}, \bibinfo {author} {\bibfnamefont {T.}~\bibnamefont
  {Vegge}}, \bibinfo {author} {\bibfnamefont {L.}~\bibnamefont {Vilhelmsen}},
  \bibinfo {author} {\bibfnamefont {M.}~\bibnamefont {Walter}}, \bibinfo
  {author} {\bibfnamefont {Z.}~\bibnamefont {Zeng}},\ and\ \bibinfo {author}
  {\bibfnamefont {K.~W.}\ \bibnamefont {Jacobsen}},\ }\bibfield  {title}
  {\enquote {\bibinfo {title} {The atomic simulation environment---a {{Python}}
  library for working with atoms},}\ }\href
  {https://doi.org/10.1088/1361-648X/aa680e} {\bibfield  {journal} {\bibinfo
  {journal} {\authorjournal{J. Phys.: Condens. Matter}{Journal of Physics:
  Condensed Matter}}\ }\textbf {\bibinfo {volume} {29}},\ \bibinfo {pages}
  {273002} (\bibinfo {year} {2017})}\BibitemShut {NoStop}%
\bibitem [{\citenamefont {Liu}, \citenamefont {He},\ and\ \citenamefont
  {Mo}(2023)}]{liuDiscrepanciesErrorEvaluation2023}%
  \BibitemOpen
  \bibfield  {author} {\bibinfo {author} {\bibfnamefont {Y.}~\bibnamefont
  {Liu}}, \bibinfo {author} {\bibfnamefont {X.}~\bibnamefont {He}},\ and\
  \bibinfo {author} {\bibfnamefont {Y.}~\bibnamefont {Mo}},\ }\bibfield
  {title} {\enquote {\bibinfo {title} {Discrepancies and error evaluation
  metrics for machine learning interatomic potentials},}\ }\href
  {https://doi.org/10.1038/s41524-023-01123-3} {\bibfield  {journal} {\bibinfo
  {journal} {\authorjournal{npj Comput. Mater.}{npj Computational Materials}}\
  }\textbf {\bibinfo {volume} {9}},\ \bibinfo {pages} {174} (\bibinfo {year}
  {2023})}\BibitemShut {NoStop}%
\bibitem [{\citenamefont {Lee}\ \emph {et~al.}(2019)\citenamefont {Lee},
  \citenamefont {Yoo}, \citenamefont {Jeong},\ and\ \citenamefont
  {Han}}]{leeSIMPLENNEfficientPackage2019}%
  \BibitemOpen
  \bibfield  {author} {\bibinfo {author} {\bibfnamefont {K.}~\bibnamefont
  {Lee}}, \bibinfo {author} {\bibfnamefont {D.}~\bibnamefont {Yoo}}, \bibinfo
  {author} {\bibfnamefont {W.}~\bibnamefont {Jeong}},\ and\ \bibinfo {author}
  {\bibfnamefont {S.}~\bibnamefont {Han}},\ }\bibfield  {title} {\enquote
  {\bibinfo {title} {{{SIMPLE-NN}}: {{An}} efficient package for training and
  executing neural-network interatomic potentials},}\ }\href
  {https://doi.org/10.1016/j.cpc.2019.04.014} {\bibfield  {journal} {\bibinfo
  {journal} {\authorjournal{Comput. Phys. Commun.}{Computer Physics
  Communications}}\ }\textbf {\bibinfo {volume} {242}},\ \bibinfo {pages}
  {95--103} (\bibinfo {year} {2019})}\BibitemShut {NoStop}%
\bibitem [{\citenamefont {Thompson}\ \emph {et~al.}(2022)\citenamefont
  {Thompson}, \citenamefont {Aktulga}, \citenamefont {Berger}, \citenamefont
  {Bolintineanu}, \citenamefont {Brown}, \citenamefont {Crozier}, \citenamefont
  {{in 't Veld}}, \citenamefont {Kohlmeyer}, \citenamefont {Moore},
  \citenamefont {Nguyen}, \citenamefont {Shan}, \citenamefont {Stevens},
  \citenamefont {Tranchida}, \citenamefont {Trott},\ and\ \citenamefont
  {Plimpton}}]{thompsonLAMMPSFlexibleSimulation2022}%
  \BibitemOpen
  \bibfield  {author} {\bibinfo {author} {\bibfnamefont {A.~P.}\ \bibnamefont
  {Thompson}}, \bibinfo {author} {\bibfnamefont {H.~M.}\ \bibnamefont
  {Aktulga}}, \bibinfo {author} {\bibfnamefont {R.}~\bibnamefont {Berger}},
  \bibinfo {author} {\bibfnamefont {D.~S.}\ \bibnamefont {Bolintineanu}},
  \bibinfo {author} {\bibfnamefont {W.~M.}\ \bibnamefont {Brown}}, \bibinfo
  {author} {\bibfnamefont {P.~S.}\ \bibnamefont {Crozier}}, \bibinfo {author}
  {\bibfnamefont {P.~J.}\ \bibnamefont {{in 't Veld}}}, \bibinfo {author}
  {\bibfnamefont {A.}~\bibnamefont {Kohlmeyer}}, \bibinfo {author}
  {\bibfnamefont {S.~G.}\ \bibnamefont {Moore}}, \bibinfo {author}
  {\bibfnamefont {T.~D.}\ \bibnamefont {Nguyen}}, \bibinfo {author}
  {\bibfnamefont {R.}~\bibnamefont {Shan}}, \bibinfo {author} {\bibfnamefont
  {M.~J.}\ \bibnamefont {Stevens}}, \bibinfo {author} {\bibfnamefont
  {J.}~\bibnamefont {Tranchida}}, \bibinfo {author} {\bibfnamefont
  {C.}~\bibnamefont {Trott}},\ and\ \bibinfo {author} {\bibfnamefont {S.~J.}\
  \bibnamefont {Plimpton}},\ }\bibfield  {title} {\enquote {\bibinfo {title}
  {{{LAMMPS}} - a flexible simulation tool for particle-based materials
  modeling at the atomic, meso, and continuum scales},}\ }\href
  {https://doi.org/10.1016/j.cpc.2021.108171} {\bibfield  {journal} {\bibinfo
  {journal} {\authorjournal{Comput. Phys. Commun.}{Computer Physics
  Communications}}\ }\textbf {\bibinfo {volume} {271}},\ \bibinfo {pages}
  {108171} (\bibinfo {year} {2022})}\BibitemShut {NoStop}%
\bibitem [{\citenamefont {Stillinger}\ and\ \citenamefont
  {Weber}(1985)}]{stillingerComputerSimulationLocal1985}%
  \BibitemOpen
  \bibfield  {author} {\bibinfo {author} {\bibfnamefont {F.~H.}\ \bibnamefont
  {Stillinger}}\ and\ \bibinfo {author} {\bibfnamefont {T.~A.}\ \bibnamefont
  {Weber}},\ }\bibfield  {title} {\enquote {\bibinfo {title} {Computer
  simulation of local order in condensed phases of silicon},}\ }\href
  {https://doi.org/10.1103/PhysRevB.31.5262} {\bibfield  {journal} {\bibinfo
  {journal} {\authorjournal{Phys. Rev. B}{Physical Review B}}\ }\textbf
  {\bibinfo {volume} {31}},\ \bibinfo {pages} {5262--5271} (\bibinfo {year}
  {1985})}\BibitemShut {NoStop}%
\bibitem [{\citenamefont {Dickel}, \citenamefont {Nitol},\ and\ \citenamefont
  {Barrett}(2021)}]{dickelLAMMPSImplementationRapid2021}%
  \BibitemOpen
  \bibfield  {author} {\bibinfo {author} {\bibfnamefont {D.}~\bibnamefont
  {Dickel}}, \bibinfo {author} {\bibfnamefont {M.}~\bibnamefont {Nitol}},\ and\
  \bibinfo {author} {\bibfnamefont {C.}~\bibnamefont {Barrett}},\ }\bibfield
  {title} {\enquote {\bibinfo {title} {{{LAMMPS}} implementation of rapid
  artificial neural network derived interatomic potentials},}\ }\href
  {https://doi.org/10.1016/j.commatsci.2021.110481} {\bibfield  {journal}
  {\bibinfo  {journal} {\authorjournal{Comput. Mater. Sci.}{Computational
  Materials Science}}\ }\textbf {\bibinfo {volume} {196}},\ \bibinfo {pages}
  {110481} (\bibinfo {year} {2021})}\BibitemShut {NoStop}%
\bibitem [{\citenamefont {Hu}\ and\ \citenamefont
  {Spain}(1984)}]{huPhasesSiliconHigh1984}%
  \BibitemOpen
  \bibfield  {author} {\bibinfo {author} {\bibfnamefont {J.}~\bibnamefont
  {Hu}}\ and\ \bibinfo {author} {\bibfnamefont {I.}~\bibnamefont {Spain}},\
  }\bibfield  {title} {\enquote {\bibinfo {title} {Phases of silicon at high
  pressure},}\ }\href {https://doi.org/10.1016/0038-1098(84)90683-5} {\bibfield
   {journal} {\bibinfo  {journal} {\authorjournal{Solid State Commun.}{Solid
  State Communications}}\ }\textbf {\bibinfo {volume} {51}},\ \bibinfo {pages}
  {263--266} (\bibinfo {year} {1984})}\BibitemShut {NoStop}%
\bibitem [{\citenamefont {McMahon}\ and\ \citenamefont
  {Nelmes}(1993)}]{mcmahonNewHighpressurePhase1993}%
  \BibitemOpen
  \bibfield  {author} {\bibinfo {author} {\bibfnamefont {M.~I.}\ \bibnamefont
  {McMahon}}\ and\ \bibinfo {author} {\bibfnamefont {R.~J.}\ \bibnamefont
  {Nelmes}},\ }\bibfield  {title} {\enquote {\bibinfo {title} {New
  high-pressure phase of {{Si}}},}\ }\href
  {https://doi.org/10.1103/PhysRevB.47.8337} {\bibfield  {journal} {\bibinfo
  {journal} {\authorjournal{Phys. Rev. B}{Physical Review B}}\ }\textbf
  {\bibinfo {volume} {47}},\ \bibinfo {pages} {8337--8340} (\bibinfo {year}
  {1993})}\BibitemShut {NoStop}%
\bibitem [{\citenamefont {Birch}(1947)}]{birchFiniteElasticStrain1947}%
  \BibitemOpen
  \bibfield  {author} {\bibinfo {author} {\bibfnamefont {F.}~\bibnamefont
  {Birch}},\ }\bibfield  {title} {\enquote {\bibinfo {title} {Finite {{Elastic
  Strain}} of {{Cubic Crystals}}},}\ }\href
  {https://doi.org/10.1103/PhysRev.71.809} {\bibfield  {journal} {\bibinfo
  {journal} {\authorjournal{Phys. Rev.}{Physical Review}}\ }\textbf {\bibinfo
  {volume} {71}},\ \bibinfo {pages} {809--824} (\bibinfo {year}
  {1947})}\BibitemShut {NoStop}%
\bibitem [{\citenamefont {Thomas}\ \emph {et~al.}(2010)\citenamefont {Thomas},
  \citenamefont {Turney}, \citenamefont {Iutzi}, \citenamefont {Amon},\ and\
  \citenamefont {McGaughey}}]{thomasPredictingPhononDispersion2010}%
  \BibitemOpen
  \bibfield  {author} {\bibinfo {author} {\bibfnamefont {J.~A.}\ \bibnamefont
  {Thomas}}, \bibinfo {author} {\bibfnamefont {J.~E.}\ \bibnamefont {Turney}},
  \bibinfo {author} {\bibfnamefont {R.~M.}\ \bibnamefont {Iutzi}}, \bibinfo
  {author} {\bibfnamefont {C.~H.}\ \bibnamefont {Amon}},\ and\ \bibinfo
  {author} {\bibfnamefont {A.~J.~H.}\ \bibnamefont {McGaughey}},\ }\bibfield
  {title} {\enquote {\bibinfo {title} {Predicting phonon dispersion relations
  and lifetimes from the spectral energy density},}\ }\href
  {https://doi.org/10.1103/PhysRevB.81.081411} {\bibfield  {journal} {\bibinfo
  {journal} {\authorjournal{Phys. Rev. B}{Physical Review B}}\ }\textbf
  {\bibinfo {volume} {81}},\ \bibinfo {pages} {081411} (\bibinfo {year}
  {2010})}\BibitemShut {NoStop}%
\bibitem [{\citenamefont {{\v S}tich}, \citenamefont {Car},\ and\ \citenamefont
  {Parrinello}(1991)}]{stichStructuralBondingDynamical1991}%
  \BibitemOpen
  \bibfield  {author} {\bibinfo {author} {\bibfnamefont {I.}~\bibnamefont {{\v
  S}tich}}, \bibinfo {author} {\bibfnamefont {R.}~\bibnamefont {Car}},\ and\
  \bibinfo {author} {\bibfnamefont {M.}~\bibnamefont {Parrinello}},\ }\bibfield
   {title} {\enquote {\bibinfo {title} {Structural, bonding, dynamical, and
  electronic properties of liquid silicon: {{An}} {\emph{ab initio}}
  molecular-dynamics study},}\ }\href
  {https://doi.org/10.1103/PhysRevB.44.4262} {\bibfield  {journal} {\bibinfo
  {journal} {\authorjournal{Phys. Rev. B}{Physical Review B}}\ }\textbf
  {\bibinfo {volume} {44}},\ \bibinfo {pages} {4262--4274} (\bibinfo {year}
  {1991})}\BibitemShut {NoStop}%
\bibitem [{\citenamefont {Custer}\ \emph {et~al.}(1994)\citenamefont {Custer},
  \citenamefont {Thompson}, \citenamefont {Jacobson}, \citenamefont {Poate},
  \citenamefont {Roorda}, \citenamefont {Sinke},\ and\ \citenamefont
  {Spaepen}}]{custerDensityAmorphousSi1994}%
  \BibitemOpen
  \bibfield  {author} {\bibinfo {author} {\bibfnamefont {J.~S.}\ \bibnamefont
  {Custer}}, \bibinfo {author} {\bibfnamefont {M.~O.}\ \bibnamefont
  {Thompson}}, \bibinfo {author} {\bibfnamefont {D.~C.}\ \bibnamefont
  {Jacobson}}, \bibinfo {author} {\bibfnamefont {J.~M.}\ \bibnamefont {Poate}},
  \bibinfo {author} {\bibfnamefont {S.}~\bibnamefont {Roorda}}, \bibinfo
  {author} {\bibfnamefont {W.~C.}\ \bibnamefont {Sinke}},\ and\ \bibinfo
  {author} {\bibfnamefont {F.}~\bibnamefont {Spaepen}},\ }\bibfield  {title}
  {\enquote {\bibinfo {title} {Density of amorphous {{Si}}},}\ }\href
  {https://doi.org/10.1063/1.111121} {\bibfield  {journal} {\bibinfo  {journal}
  {\authorjournal{Appl. Phys. Lett.}{Applied Physics Letters}}\ }\textbf
  {\bibinfo {volume} {64}},\ \bibinfo {pages} {437--439} (\bibinfo {year}
  {1994})}\BibitemShut {NoStop}%
\bibitem [{\citenamefont {Santos}\ \emph {et~al.}(2019)\citenamefont {Santos},
  \citenamefont {Aboy}, \citenamefont {Marqu{\'e}s}, \citenamefont
  {L{\'o}pez},\ and\ \citenamefont {Pelaz}}]{santosGenerationAmorphousSi2019}%
  \BibitemOpen
  \bibfield  {author} {\bibinfo {author} {\bibfnamefont {I.}~\bibnamefont
  {Santos}}, \bibinfo {author} {\bibfnamefont {M.}~\bibnamefont {Aboy}},
  \bibinfo {author} {\bibfnamefont {L.~A.}\ \bibnamefont {Marqu{\'e}s}},
  \bibinfo {author} {\bibfnamefont {P.}~\bibnamefont {L{\'o}pez}},\ and\
  \bibinfo {author} {\bibfnamefont {L.}~\bibnamefont {Pelaz}},\ }\bibfield
  {title} {\enquote {\bibinfo {title} {Generation of amorphous {{Si}}
  structurally compatible with experimental samples through the quenching
  process: {{A}} systematic molecular dynamics simulation study},}\ }\href
  {https://doi.org/10.1016/j.jnoncrysol.2018.09.024} {\bibfield  {journal}
  {\bibinfo  {journal} {\authorjournal{J. Non-Cryst. Solids}{Journal of
  Non-Crystalline Solids}}\ }\textbf {\bibinfo {volume} {503--504}},\ \bibinfo
  {pages} {20--27} (\bibinfo {year} {2019})}\BibitemShut {NoStop}%
\bibitem [{\citenamefont {Deringer}\ \emph {et~al.}(2018)\citenamefont
  {Deringer}, \citenamefont {Bernstein}, \citenamefont {Bart{\'o}k},
  \citenamefont {Cliffe}, \citenamefont {Kerber}, \citenamefont {Marbella},
  \citenamefont {Grey}, \citenamefont {Elliott},\ and\ \citenamefont
  {Cs{\'a}nyi}}]{deringerRealisticAtomisticStructure2018}%
  \BibitemOpen
  \bibfield  {author} {\bibinfo {author} {\bibfnamefont {V.~L.}\ \bibnamefont
  {Deringer}}, \bibinfo {author} {\bibfnamefont {N.}~\bibnamefont {Bernstein}},
  \bibinfo {author} {\bibfnamefont {A.~P.}\ \bibnamefont {Bart{\'o}k}},
  \bibinfo {author} {\bibfnamefont {M.~J.}\ \bibnamefont {Cliffe}}, \bibinfo
  {author} {\bibfnamefont {R.~N.}\ \bibnamefont {Kerber}}, \bibinfo {author}
  {\bibfnamefont {L.~E.}\ \bibnamefont {Marbella}}, \bibinfo {author}
  {\bibfnamefont {C.~P.}\ \bibnamefont {Grey}}, \bibinfo {author}
  {\bibfnamefont {S.~R.}\ \bibnamefont {Elliott}},\ and\ \bibinfo {author}
  {\bibfnamefont {G.}~\bibnamefont {Cs{\'a}nyi}},\ }\bibfield  {title}
  {\enquote {\bibinfo {title} {Realistic {{Atomistic Structure}} of {{Amorphous
  Silicon}} from {{Machine-Learning-Driven Molecular Dynamics}}},}\ }\href
  {https://doi.org/10.1021/acs.jpclett.8b00902} {\bibfield  {journal} {\bibinfo
   {journal} {\authorjournal{J. Phys. Chem. Lett.}{The Journal of Physical
  Chemistry Letters}}\ }\textbf {\bibinfo {volume} {9}},\ \bibinfo {pages}
  {2879--2885} (\bibinfo {year} {2018})}\BibitemShut {NoStop}%
\bibitem [{\citenamefont {Wolkow}(1992)}]{wolkowDirectObservationIncrease1992}%
  \BibitemOpen
  \bibfield  {author} {\bibinfo {author} {\bibfnamefont {R.~A.}\ \bibnamefont
  {Wolkow}},\ }\bibfield  {title} {\enquote {\bibinfo {title} {Direct
  observation of an increase in buckled dimers on {{Si}}(001) at low
  temperature},}\ }\href {https://doi.org/10.1103/PhysRevLett.68.2636}
  {\bibfield  {journal} {\bibinfo  {journal} {\authorjournal{Phys. Rev.
  Lett.}{Physical Review Letters}}\ }\textbf {\bibinfo {volume} {68}},\
  \bibinfo {pages} {2636--2639} (\bibinfo {year} {1992})}\BibitemShut {NoStop}%
\bibitem [{\citenamefont {Xu}\ \emph {et~al.}(2004)\citenamefont {Xu},
  \citenamefont {Deng}, \citenamefont {Yu}, \citenamefont {Tong}, \citenamefont
  {Van~Hove}, \citenamefont {Jona},\ and\ \citenamefont
  {Zasada}}]{xuAtomicStructureCleaved2004}%
  \BibitemOpen
  \bibfield  {author} {\bibinfo {author} {\bibfnamefont {G.}~\bibnamefont
  {Xu}}, \bibinfo {author} {\bibfnamefont {B.}~\bibnamefont {Deng}}, \bibinfo
  {author} {\bibfnamefont {Z.}~\bibnamefont {Yu}}, \bibinfo {author}
  {\bibfnamefont {S.~Y.}\ \bibnamefont {Tong}}, \bibinfo {author}
  {\bibfnamefont {M.~A.}\ \bibnamefont {Van~Hove}}, \bibinfo {author}
  {\bibfnamefont {F.}~\bibnamefont {Jona}},\ and\ \bibinfo {author}
  {\bibfnamefont {I.}~\bibnamefont {Zasada}},\ }\bibfield  {title} {\enquote
  {\bibinfo {title} {Atomic structure of the cleaved {{Si}}(111)-(2\texttimes
  1) surface refined by dynamical {{LEED}}},}\ }\href
  {https://doi.org/10.1103/PhysRevB.70.045307} {\bibfield  {journal} {\bibinfo
  {journal} {\authorjournal{Phys. Rev. B}{Physical Review B}}\ }\textbf
  {\bibinfo {volume} {70}},\ \bibinfo {pages} {045307} (\bibinfo {year}
  {2004})}\BibitemShut {NoStop}%
\bibitem [{\citenamefont {Zitzlsperger}\ \emph {et~al.}(1997)\citenamefont
  {Zitzlsperger}, \citenamefont {Honke}, \citenamefont {Pavone},\ and\
  \citenamefont {Schr{\"o}der}}]{zitzlspergerInitioCalculationStructure1997a}%
  \BibitemOpen
  \bibfield  {author} {\bibinfo {author} {\bibfnamefont {M.}~\bibnamefont
  {Zitzlsperger}}, \bibinfo {author} {\bibfnamefont {R.}~\bibnamefont {Honke}},
  \bibinfo {author} {\bibfnamefont {P.}~\bibnamefont {Pavone}},\ and\ \bibinfo
  {author} {\bibfnamefont {U.}~\bibnamefont {Schr{\"o}der}},\ }\bibfield
  {title} {\enquote {\bibinfo {title} {Ab initio calculation of the structure,
  the electronic states and the phonon dispersion of the
  {{Si}}(111)-(2\texttimes 1) surface},}\ }\href
  {https://doi.org/10.1016/S0039-6028(96)01373-8} {\bibfield  {journal}
  {\bibinfo  {journal} {\authorjournal{Surf. Sci.}{Surface Science}}\ }\textbf
  {\bibinfo {volume} {377--379}},\ \bibinfo {pages} {108--111} (\bibinfo {year}
  {1997})}\BibitemShut {NoStop}%
\bibitem [{\citenamefont {Meade}\ and\ \citenamefont
  {Vanderbilt}(1989)}]{meadeAdatomsSi111Ge1111989}%
  \BibitemOpen
  \bibfield  {author} {\bibinfo {author} {\bibfnamefont {R.~D.}\ \bibnamefont
  {Meade}}\ and\ \bibinfo {author} {\bibfnamefont {D.}~\bibnamefont
  {Vanderbilt}},\ }\bibfield  {title} {\enquote {\bibinfo {title} {Adatoms on
  {{Si}}(111) and {{Ge}}(111) surfaces},}\ }\href
  {https://doi.org/10.1103/PhysRevB.40.3905} {\bibfield  {journal} {\bibinfo
  {journal} {\authorjournal{Phys. Rev. B}{Physical Review B}}\ }\textbf
  {\bibinfo {volume} {40}},\ \bibinfo {pages} {3905--3913} (\bibinfo {year}
  {1989})}\BibitemShut {NoStop}%
\bibitem [{\citenamefont {Solares}\ \emph {et~al.}(2005)\citenamefont
  {Solares}, \citenamefont {Dasgupta}, \citenamefont {Schultz}, \citenamefont
  {Kim}, \citenamefont {Musgrave},\ and\ \citenamefont
  {Goddard}}]{solaresDensityFunctionalTheory2005}%
  \BibitemOpen
  \bibfield  {author} {\bibinfo {author} {\bibfnamefont {S.~D.}\ \bibnamefont
  {Solares}}, \bibinfo {author} {\bibfnamefont {S.}~\bibnamefont {Dasgupta}},
  \bibinfo {author} {\bibfnamefont {P.~A.}\ \bibnamefont {Schultz}}, \bibinfo
  {author} {\bibfnamefont {Y.-H.}\ \bibnamefont {Kim}}, \bibinfo {author}
  {\bibfnamefont {C.~B.}\ \bibnamefont {Musgrave}},\ and\ \bibinfo {author}
  {\bibfnamefont {W.~A.}\ \bibnamefont {Goddard}},\ }\bibfield  {title}
  {\enquote {\bibinfo {title} {Density {{Functional Theory Study}} of the
  {{Geometry}}, {{Energetics}}, and {{Reconstruction Process}} of {{Si}}(111)
  {{Surfaces}}},}\ }\href {https://doi.org/10.1021/la052029s} {\bibfield
  {journal} {\bibinfo  {journal} {Langmuir}\ }\textbf {\bibinfo {volume}
  {21}},\ \bibinfo {pages} {12404--12414} (\bibinfo {year} {2005})}\BibitemShut
  {NoStop}%
\bibitem [{\citenamefont {Takayanagi}\ \emph {et~al.}(1985)\citenamefont
  {Takayanagi}, \citenamefont {Tanishiro}, \citenamefont {Takahashi},\ and\
  \citenamefont {Takahashi}}]{takayanagiStructureAnalysisSi11171985}%
  \BibitemOpen
  \bibfield  {author} {\bibinfo {author} {\bibfnamefont {K.}~\bibnamefont
  {Takayanagi}}, \bibinfo {author} {\bibfnamefont {Y.}~\bibnamefont
  {Tanishiro}}, \bibinfo {author} {\bibfnamefont {S.}~\bibnamefont
  {Takahashi}},\ and\ \bibinfo {author} {\bibfnamefont {M.}~\bibnamefont
  {Takahashi}},\ }\bibfield  {title} {\enquote {\bibinfo {title} {Structure
  analysis of {{Si}}(111)-7\texttimes 7 reconstructed surface by transmission
  electron diffraction},}\ }\href
  {https://doi.org/10.1016/0039-6028(85)90753-8} {\bibfield  {journal}
  {\bibinfo  {journal} {\authorjournal{Surf. Sci.}{Surface Science}}\ }\textbf
  {\bibinfo {volume} {164}},\ \bibinfo {pages} {367--392} (\bibinfo {year}
  {1985})}\BibitemShut {NoStop}%
\bibitem [{\citenamefont {Li}\ \emph {et~al.}(1988)\citenamefont {Li},
  \citenamefont {Chen}, \citenamefont {Allen},\ and\ \citenamefont
  {Broughton}}]{liEnergyVibrationalSpectrum1988a}%
  \BibitemOpen
  \bibfield  {author} {\bibinfo {author} {\bibfnamefont {X.-P.}\ \bibnamefont
  {Li}}, \bibinfo {author} {\bibfnamefont {G.}~\bibnamefont {Chen}}, \bibinfo
  {author} {\bibfnamefont {P.~B.}\ \bibnamefont {Allen}},\ and\ \bibinfo
  {author} {\bibfnamefont {J.~Q.}\ \bibnamefont {Broughton}},\ }\bibfield
  {title} {\enquote {\bibinfo {title} {Energy and vibrational spectrum of the
  {{Si}}(111) (7\texttimes 7) surface from empirical potentials},}\ }\href
  {https://doi.org/10.1103/PhysRevB.38.3331} {\bibfield  {journal} {\bibinfo
  {journal} {\authorjournal{Phys. Rev. B}{Physical Review B}}\ }\textbf
  {\bibinfo {volume} {38}},\ \bibinfo {pages} {3331--3341} (\bibinfo {year}
  {1988})}\BibitemShut {NoStop}%
\bibitem [{\citenamefont {Nishimura}\ \emph {et~al.}(2026)\citenamefont
  {Nishimura}, \citenamefont {Kubota}, \citenamefont {Takematsu}, \citenamefont
  {Hirai},\ and\ \citenamefont {Watanabe}}]{nishimuraTersoffNNZenodo2026}%
  \BibitemOpen
  \bibfield  {author} {\bibinfo {author} {\bibfnamefont {Y.}~\bibnamefont
  {Nishimura}}, \bibinfo {author} {\bibfnamefont {S.}~\bibnamefont {Kubota}},
  \bibinfo {author} {\bibfnamefont {K.}~\bibnamefont {Takematsu}}, \bibinfo
  {author} {\bibfnamefont {K.}~\bibnamefont {Hirai}},\ and\ \bibinfo {author}
  {\bibfnamefont {T.}~\bibnamefont {Watanabe}},\ }\href
  {https://doi.org/10.5281/zenodo.23290766} {\enquote {\bibinfo {title}
  {{Tersoff-NN --- Software, Trained Models, and Reference Dataset for
  Silicon}},}\ }\bibinfo {howpublished} {Zenodo} (\bibinfo {year} {2026}),\
  \bibinfo {note} {https://doi.org/10.5281/zenodo.23290766}\BibitemShut
  {NoStop}%
\end{thebibliography}%

\end{document}